\documentclass[twocolumn]{aastex631}

\usepackage{amsmath}
\usepackage{mathtools}
\usepackage{xspace}
\usepackage{CJKutf8}

\newcommand{\ngal}{12}

\newcommand{\redrockets}{UFOs} 
\newcommand{\parrot}{\texttt{parrot}}
\newcommand{\eazy}{\texttt{EAzY}}
\newcommand{\prospector}{\texttt{Prospector}}
\newcommand{\mstar}{M$_*$}

\newcommand{\JWST}{\emph{JWST}\xspace}
\newcommand{\HST}{\emph{HST}\xspace}

\newcommand{\re}{\ensuremath{R_{\mathrm{e}}}\,}

\begin{document}


\title{\JWST reveals a population of ultra-red, flattened disk galaxies at $2\lesssim z \lesssim 6$ previously missed by \HST}

\author[0000-0002-7524-374X]{Erica J. Nelson}
\affiliation{Department for Astrophysical and Planetary Science, University of Colorado, Boulder, CO 80309, USA}

\author[0000-0002-1714-1905]{Katherine A. Suess}
\affiliation{Department of Astronomy and Astrophysics, University of California, Santa Cruz, 1156 High Street, Santa Cruz, CA 95064 USA}
\affiliation{Kavli Institute for Particle Astrophysics and Cosmology and Department of Physics, Stanford University, Stanford, CA 94305, USA}

\author[0000-0001-5063-8254]{Rachel Bezanson}
\affiliation{Department of Physics and Astronomy and PITT PACC, University of Pittsburgh, Pittsburgh, PA 15260, USA}

\author[0000-0002-0108-4176]{Sedona H. Price}
\affiliation{Max-Planck-Institut f\"{u}r extraterrestrische Physik (MPE), Giessenbachstr. 1, D-85748 Garching, Germany}

\author[0000-0002-8282-9888]{Pieter van Dokkum}
\affiliation{Astronomy Department, Yale University, 52 Hillhouse Ave,
New Haven, CT 06511, USA}

\author[0000-0001-6755-1315]{Joel Leja}
\affiliation{Department of Astronomy \& Astrophysics, The Pennsylvania
State University, University Park, PA 16802, USA}
\affiliation{Institute for Computational \& Data Sciences, The Pennsylvania State University, University Park, PA, USA}
\affiliation{Institute for Gravitation and the Cosmos, The Pennsylvania State University, University Park, PA 16802, USA}

\author[0000-0001-9269-5046]{Bingjie Wang (\begin{CJK*}{UTF8}{gbsn}王冰洁\ignorespacesafterend\end{CJK*})}
\affiliation{Department of Astronomy \& Astrophysics, The Pennsylvania
State University, University Park, PA 16802, USA}
\affiliation{Institute for Computational \& Data Sciences, The Pennsylvania State University, University Park, PA, USA}
\affiliation{Institute for Gravitation and the Cosmos, The Pennsylvania State University, University Park, PA 16802, USA}

\author[0000-0001-7160-3632]{Katherine E. Whitaker}
\affil{Department of Astronomy, University of Massachusetts, Amherst, MA 01003, USA}
\affiliation{Cosmic Dawn Center (DAWN), Denmark}

\author[0000-0002-2057-5376]{Ivo Labb\'{e}} 
\affiliation{Centre for Astrophysics and Supercomputing, Swinburne University of Technology, Melbourne, VIC 3122, Australia}

\author[0000-0003-1641-6185]{Laia Barrufet}
\affiliation{Department of Astronomy, University of Geneva, Chemin Pegasi 51, 1290 Versoix, Switzerland}

\author[0000-0003-2680-005X]{Gabriel Brammer}
\affiliation{Cosmic Dawn Center (DAWN), Niels Bohr Institute, University of Copenhagen, Jagtvej 128, K\o benhavn N, DK-2200, Denmark}

\author[0000-0002-2929-3121]{Daniel J. Eisenstein}
\affiliation{Center for Astrophysics $|$ Harvard \& Smithsonian, 60 Garden Street, Cambridge, MA 02138, USA}

\author[0000-0002-9280-7594]{Benjamin~D.~Johnson}
\affiliation{Center for Astrophysics $|$ Harvard \& Smithsonian, 60 Garden Street, Cambridge, MA 02138, USA}

\author[0000-0002-9389-7413]{Kasper E. Heintz}
\affiliation{Cosmic Dawn Center (DAWN), Niels Bohr Institute, University of Copenhagen, Jagtvej 128, K\o benhavn N, DK-2200, Denmark}

\author[0000-0003-0384-0681]{Elijah Mathews}
\affiliation{Department of Astronomy \& Astrophysics, The Pennsylvania
State University, University Park, PA 16802, USA}
\affiliation{Institute for Computational \& Data Sciences, The Pennsylvania State University, University Park, PA, USA}
\affiliation{Institute for Gravitation and the Cosmos, The Pennsylvania State University, University Park, PA 16802, USA}

\author[0000-0001-8367-6265]{Tim B. Miller}
\affiliation{Astronomy Department, Yale University, 52 Hillhouse Ave,
New Haven, CT 06511, USA}

\author[0000-0001-5851-6649]{Pascal A. Oesch}
\affiliation{Department of Astronomy, University of Geneva, Chemin Pegasi 51, 1290 Versoix, Switzerland}
\affiliation{Cosmic Dawn Center (DAWN), Niels Bohr Institute, University of Copenhagen, Jagtvej 128, K\o benhavn N, DK-2200, Denmark}

\author[0000-0001-9276-7062]{Lester Sandles}
\affiliation{Kavli Institute for Cosmology, University of Cambridge, Madingley Road, Cambridge, CB3 0HA, UK}
\affiliation{Cavendish Laboratory, University of Cambridge, 19 JJ Thomson Avenue, Cambridge, CB3 0HE, UK}

\author[0000-0003-4075-7393]{David J. Setton}
\affiliation{Department of Physics and Astronomy and PITT PACC, University of Pittsburgh, Pittsburgh, PA 15260, USA}

\author[0000-0003-2573-9832]{Joshua S. Speagle (\begin{CJK*}{UTF8}{gbsn}沈佳士\ignorespacesafterend\end{CJK*})}
\affiliation{David A. Dunlap Department of Astronomy \& Astrophysics, University of Toronto, 50 St George Street, Toronto ON M5S 3H4, Canada}
\affiliation{Dunlap Institute for Astronomy \& Astrophysics, University of Toronto, 50 St George Street, Toronto, ON M5S 3H4, Canada}
\affiliation{Department of Statistical Sciences, University of Toronto, 100 St George St, Toronto, ON M5S 3G3, Canada}

\author[0000-0002-8224-4505]{Sandro Tacchella}
\affiliation{Kavli Institute for Cosmology, University of Cambridge, Madingley Road, Cambridge, CB3 0HA, UK}
\affiliation{Cavendish Laboratory, University of Cambridge, 19 JJ Thomson Avenue, Cambridge, CB3 0HE, UK}

\author[0000-0001-9728-8909]{Ken-ichi Tadaki}
\affiliation{National Astronomical Observatory of Japan, 2-21-1 Osawa, Mitaka, Tokyo 181-8588, Japan}

\author[0000-0003-1614-196X]{John.~R.~Weaver}
\affiliation{Department of Astronomy, University of Massachusetts, Amherst, MA 01003, USA}

\author[0000-0003-4891-0794]{Hannah \"Ubler}
\affiliation{Kavli Institute for Cosmology, University of Cambridge, Madingley Road, Cambridge, CB3 0HA, UK}
\affiliation{Cavendish Laboratory, University of Cambridge, 19 JJ Thomson Avenue, Cambridge, CB3 0HE, UK}

\begin{abstract}

With just a month of data, \JWST is already transforming our view of the Universe, revealing and resolving starlight in unprecedented populations of galaxies. Although ``\HST-dark" galaxies have previously been detected at long wavelengths, these observations generally suffer from a lack of spatial resolution which limits our ability to characterize their sizes and morphologies. 
Here we report 
on a first view of starlight from a subset of the \HST-dark population that are bright with 
\JWST/NIRCam ($4.4\micron<24.5\mathrm{mag}$) and very faint or even invisible with 
\HST ($<1.6$\micron). In this Letter we focus on a dramatic and unanticipated population of physically extended galaxies ($\gtrsim$0.17'').
These 12 galaxies have photometric redshifts $2<z<6$, high stellar masses $M_{\star}\gtrsim 10^{10}~M_{\odot}$, and significant dust-attenuated star formation.
Surprisingly, the galaxies have elongated projected axis ratios at 4.4\micron, suggesting that the population is disk-dominated or prolate. 
Most of the galaxies appear red at all radii, suggesting significant dust attenuation throughout. We refer to these red, disky, \HST-dark galaxies as Ultra-red Flattened Objects. With $\re(\mathrm{F444W})\sim1-2$~kpc, the galaxies are similar in size to compact massive galaxies at $z\sim2$ and the cores of massive galaxies and S0s at $z\sim0$. 
The stellar masses, sizes, and morphologies of the sample suggest that some could be progenitors of lenticular or fast-rotating galaxies in the local Universe. The existence of this population suggests that our previous censuses of the universe may have missed massive, dusty edge-on disks, in addition to dust-obscured starbursts.

\end{abstract}

\keywords{galaxies: evolution -- galaxies: formation -- galaxies: high-redshift -- galaxies: structure}

\section{Introduction} \label{sec:intro}

It is well established that our census of galaxies as viewed by the {\it Hubble Space Telescope} (\emph{HST}) in rest-frame optical wavelengths is incomplete. Measurements at far-infrared and submillimeter wavelengths reveal a population of dusty star-forming galaxies that host such extreme starbursts that they are often entirely obscured by dust at ultraviolet and optical wavelengths \citep[for a review, see][]{casey:14}. This population is not negligible: in fact, they dominate the total star formation rate budget of the universe at $z\lesssim4$ \citep[e.g.,][]{madau:14,zavala:21}. Because these dusty galaxies are likely to be massive \citep[e.g.,][]{whitaker:17}, and may also be the progenitors of today's large elliptical galaxies \citep[e.g.,][]{toft:14}, our lack of ability to study their stellar properties with \emph{HST} indicates that we do not yet fully understand the growth of the most massive galaxies at cosmic noon.

To date, detailed studies of the morphologies of these dust-obscured galaxies have been challenging.  Observations with resolutions $\ll 1"$ are required in order to spatially resolve them; however, because these galaxies are only bright at long wavelengths, this requires high-resolution submillimeter/radio interferometry. Due to the limited field of view of the Atacama Large Millimeter/submillimeter Array (\emph{ALMA}) interferometer, these studies have generally been restricted to samples of galaxies already known with \HST \citep[e.g.,][]{tadaki:20} or very bright galaxies \citep[e.g.,][]{franco:18,gomez-guijarro:22,walter:16}. These studies have generally found compact far-infrared sizes \citep{hodge:16,Oteo:16,barro:16,rujopakarn:16,fujimoto:17,tadaki:17,tadaki:20,Nelson:19,gullberg:19}.
The compactness of these galaxies has been interpreted as watching these galaxies in the process of building their bulges \citep[e.g.][]{tadaki:20,hodge:20}.

While dusty star-forming galaxies have long been detectable at far-infrared wavelengths \citep[e.g.,][]{barger:98,hughes:98, coppin:06, elbaz:11} and spectroscopically confirmed to reside at high redshift \citep[e.g.,][]{chapman:05,pope:08}, the $>3$'' resolution of far-infrared instruments has made the study of the sizes and morphologies of dusty star-forming galaxies in the infrared impossible. Moreover, these known submillimeter sources are often barely detectable, if at all, by \HST in the near-infrared \citep[e.g.,][]{smail:98}. In recent years, ``\HST-dark'' and other similar terms are commonly adopted to describe extremely dusty high redshift galaxies \citep[e.g.,][]{wang:16,franco:18,williams:19, sun:21, manning:22}.  The successful launch and commissioning of the \emph{James Webb Space Telescope} (\emph{JWST}) has changed the game: now, 
sub-arcsecond observations of these galaxies are possible out to 28\,$\mu$m.  

In this paper, we report the discovery of 
29 
\HST-dark galaxies, \ngal\ of which are extended, that are fully obscured at optical wavelengths but bright at $4.4\mu$m with \emph{JWST} imaging from the Cosmic Evolution Early Release Science (CEERS) Survey.  This $4.4\,\mu$m imaging traces rest-frame wavelengths of $1.1\,\mu$m ($0.6\,\mu$m) at $z=3$ ($z=6$). 
These extended red star-forming galaxies are 
the main topic of this paper, which is organized as follows.
Section~\ref{sec:data} describes the observations and sample selection. In Section~\ref{subsec:ReSt}, we present our findings on the physical properties of this galaxy sample, including their redshifts and stellar populations.
Section~\ref{subsec:sizes} contains our identification of the shapes and sizes of these galaxies, and we set forth our results on their observed color gradients in Section~\ref{sec:color}. We discuss the findings and conclusions of this study in Section~\ref{sec:conc}.

In this paper, we assume the WMAP9 $\mathrm{\Lambda CDM}$ cosmology with $\Omega _{M}=0.2865$, $\Omega _{\Lambda}=0.7135$ and $H_0 = 69.32 \, \mathrm{km}~\mathrm{s}^{-1}~\mathrm{Mpc}^{-1}$ \citep{wmap9}.
All magnitudes in this paper are expressed in the AB system \citep{oke:74}.



\begin{figure*}
    \centering
    \includegraphics[width=\textwidth]{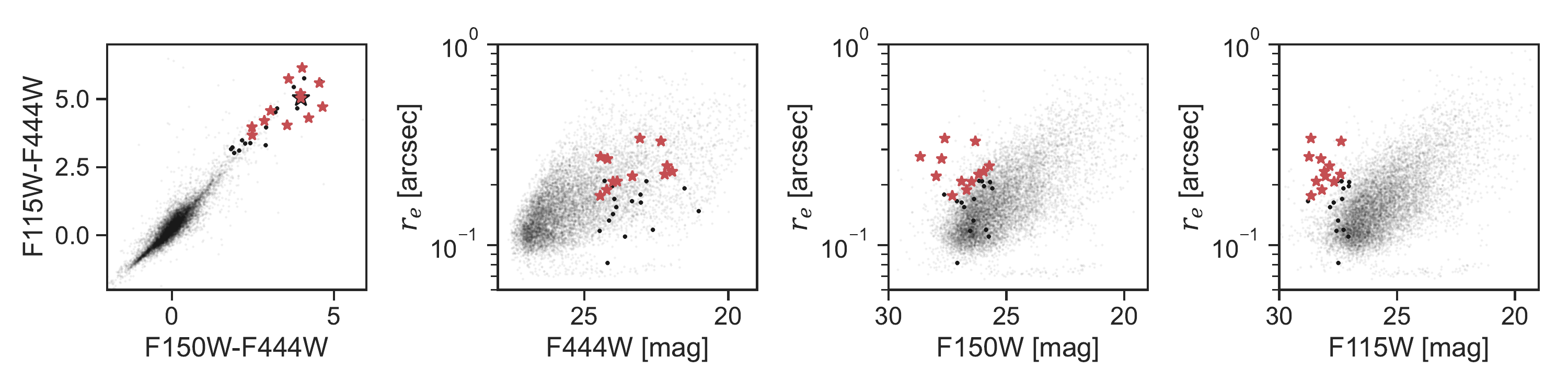}
    \includegraphics[width=.8\textwidth]{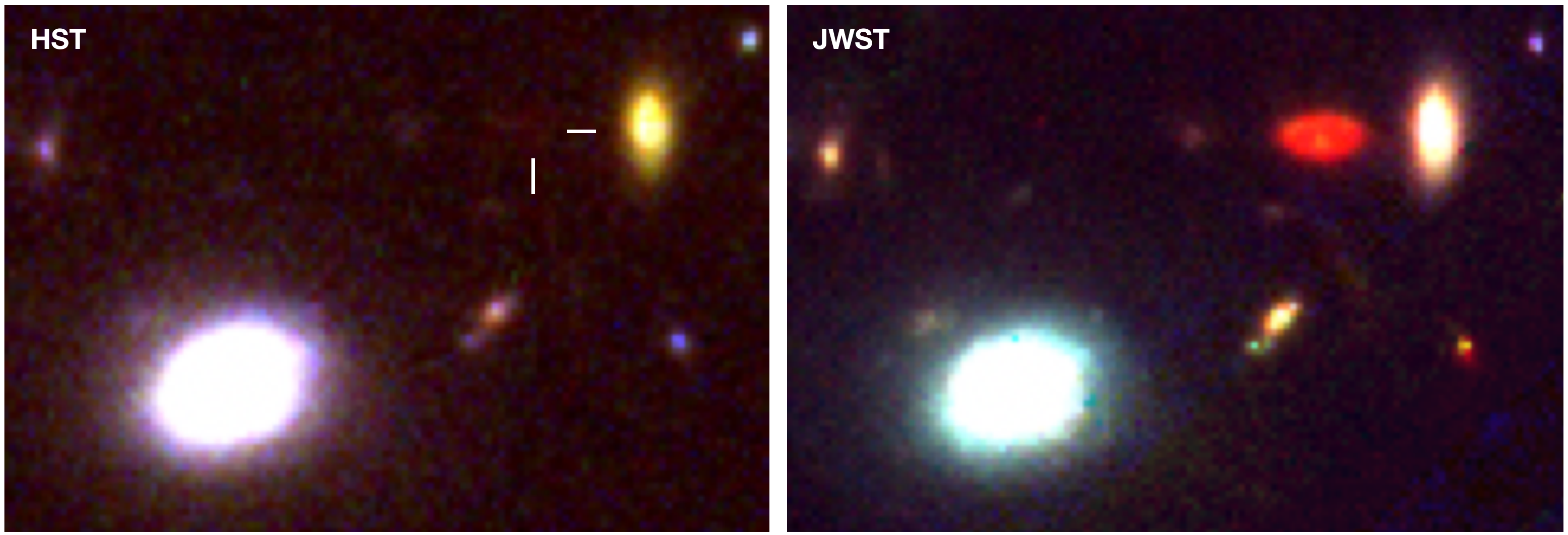}
    \caption{(Top row): A sample of 29 galaxies (black points and red stars) are selected to be bright at the reddest wavelengths \JWST/NIRCam can detect (4.4\micron), and not detected at the wavelengths previously visible with \HST, namely $<1.6\micron$ (black points). Finally, we isolate the 12 most extended galaxies with a cut on $\re > 0.17''$ (extended galaxies indicated by red stars). (Bottom row): An example color image of one extended galaxy (black star in top left panel) to demonstrate that while we do not see these objects in \HST/WFC3 imaging (bottom left: color composite of F606W, F125W, and F160W), these galaxies are extremely red and prominent in \JWST/NIRCam imaging (bottom right: color composite of F150W, F277W, F444W).}
    \label{fig:selection_v2}
\end{figure*}

\begin{figure*}
    \centering
    \includegraphics[width=0.9\textwidth]{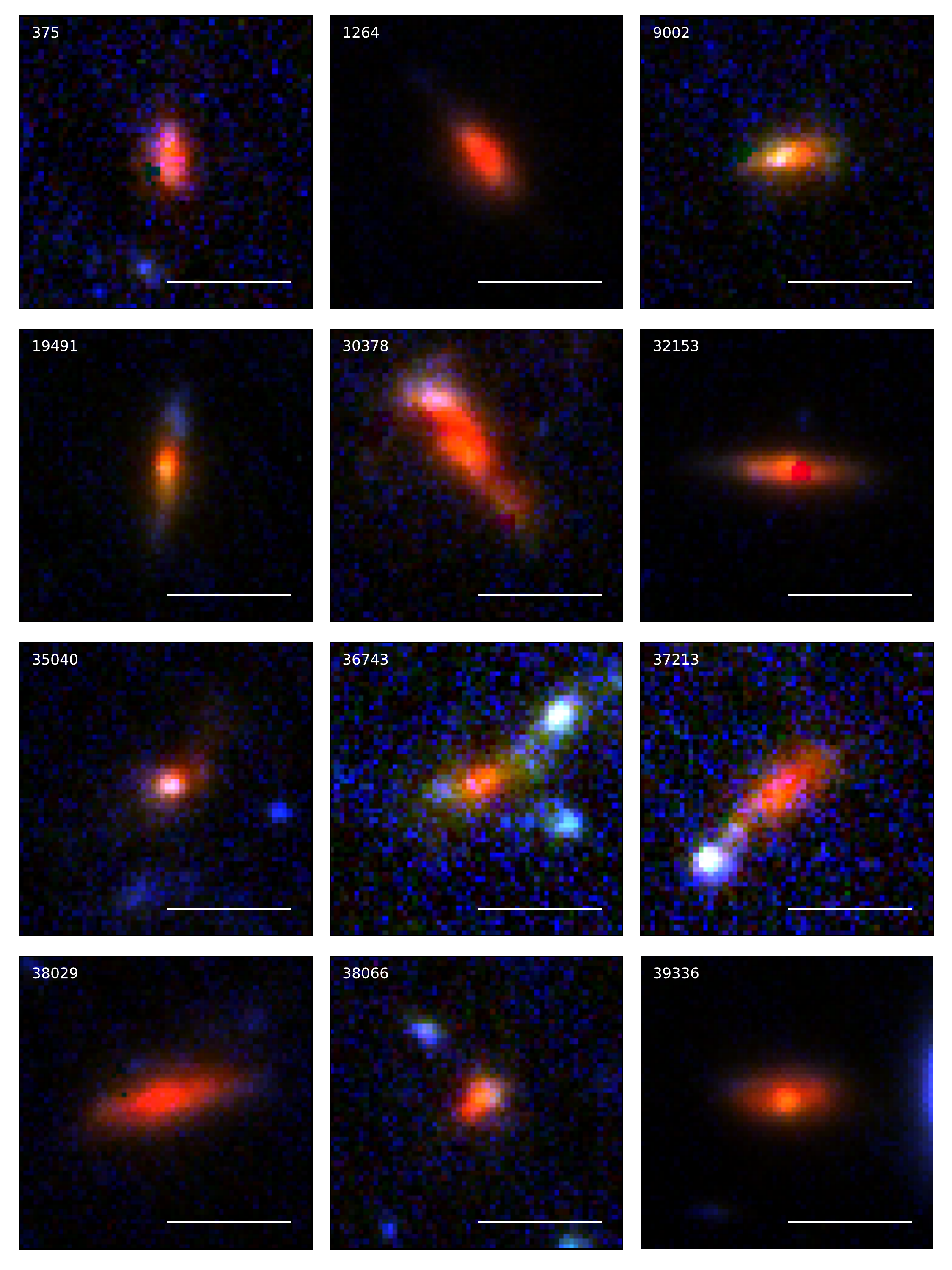}
    \caption{Three color images for all of the \redrockets\ in F150W, F277W, and F444W filters, with a 1" bar indicated. Most of these galaxies are red throughout and consistently elongated.}
    \label{fig:gallery}
\end{figure*}

\section{Data \& Sample Selection}
\label{sec:data}
For this work, we use overlapping imaging in the AEGIS field from  \emph{JWST} via the CEERS program \citep{Finkelstein:22} and \emph{HST} via the CANDELS program \citep{grogin:11,koekemoer:11}. The early CEERS imaging covers a $\sim$40 sq. arcmin portion of the AEGIS field and was taken in six broadband near-infrared filters (F115W, F150W, F200W, F277W, F356W, and F444W) and one medium band filter (F410M). Stage 2 of the \JWST calibration pipeline (v1.5.2) produced flux-calibrated exposures, publicly available from the MAST archive. Further reduction, aligning, and co-adding of the exposures was conducted using the public software package \texttt{grizli} \citep{Brammer:21} and described fully in Brammer et al. (in prep). Briefly, \texttt{grizli} masks imaging artifacts, subtracts an overall sky background, aligns the images to stars from the Gaia DR3 catalog, and projects images to a common pixel space using \texttt{astrodrizzle}. Additional background structure was removed with a 5\arcsec\ median filter after masking bright sources \citep[see also e.g.,][]{labbe:22}.

We detect and segment sources on an inverse variance weighted combination of the F277W, F356W, and F444W images convolved with a 2.5 pixel FWHM Gaussian using standard \texttt{astropy} and \texttt{photutils} procedures. Photometry is done on all detected sources in 0.32\arcsec\ and 0.5\arcsec\ diameter circular apertures. These aperture fluxes are then scaled to total fluxes using the Kron autoscaling aperture measurement from the detection image plus a small additional correction based on the encircled energy from WebbPSF \citep{perrin:15}. For additional details see \citet{labbe:22}. 

When comparing between the \JWST and \HST color images, we quickly noticed a set of large \JWST-bright galaxies that were invisible with \HST.  To collect this population, we select galaxies that are bright at the reddest wavelengths \JWST/NIRCam can detect and not detected at the $<1.6\micron$ wavelengths previously visible with \HST. Quantitatively, we identify \HST-dark galaxies with total AB magnitudes of:
\begin{enumerate}
\item $\mathrm{F444W} < 24.5~\mathrm{mag}$
\vspace{-0.2cm}
\item $\mathrm{F150W} > 25.5~\mathrm{mag}$ 
\vspace{-0.2cm}
\item $\mathrm{F115W} > 27~\mathrm{mag}$
\end{enumerate}

This selection identifies 29 \HST-dark galaxies \citep[similar to][]{barrufet:22}, with a range of projected sizes, spanning 0.08-0.33'' (see Fig.~\ref{fig:selection_v2}). Visual inspection of the F150W-F444W red objects revealed a striking population of extended galaxies (see Fig.~\ref{fig:gallery}). While these galaxies have typical sizes for their mass and redshift, they are larger than would be expected for extremely high redshift ($z\gtrsim 6$) galaxies \citep[e.g.,][]{Holwerda15,Kawamata18,naidu:22,Yang22}. We fit all galaxies that fall in our color selection with \texttt{GALFIT} \citep{peng:02, peng:10} and focus on the most extended, applying a threshold of 0.17'' to the \texttt{GALFIT} 
F444W 
half light radii.
Hence, in this paper, we highlight this previously unseen population of dusty galaxies at $2<z<6$.  A description of the properties of the full sample of \HST-dark galaxies is presented in \cite{barrufet:22}.
These requirements result in the sample of \ngal\ galaxies shown in Fig.~\ref{fig:gallery}. The selection is shown in Fig.~\ref{fig:selection_v2}. These galaxies constitute some of the reddest, brightest sources, all undetected by \HST.

\begin{figure*}
    \centering
\includegraphics[ clip,width=0.3\textwidth]{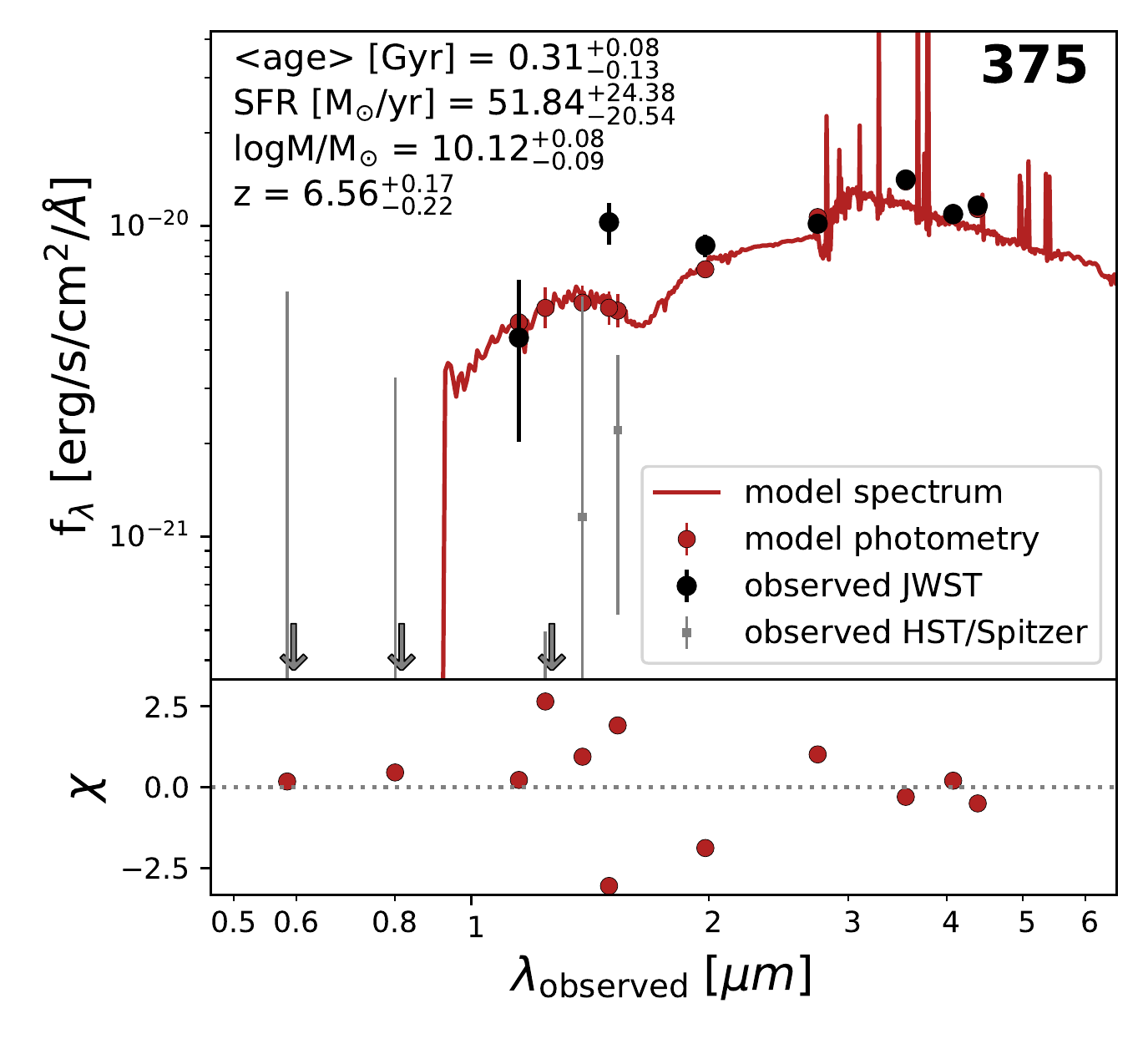}
\includegraphics[ clip,width=0.3\textwidth]{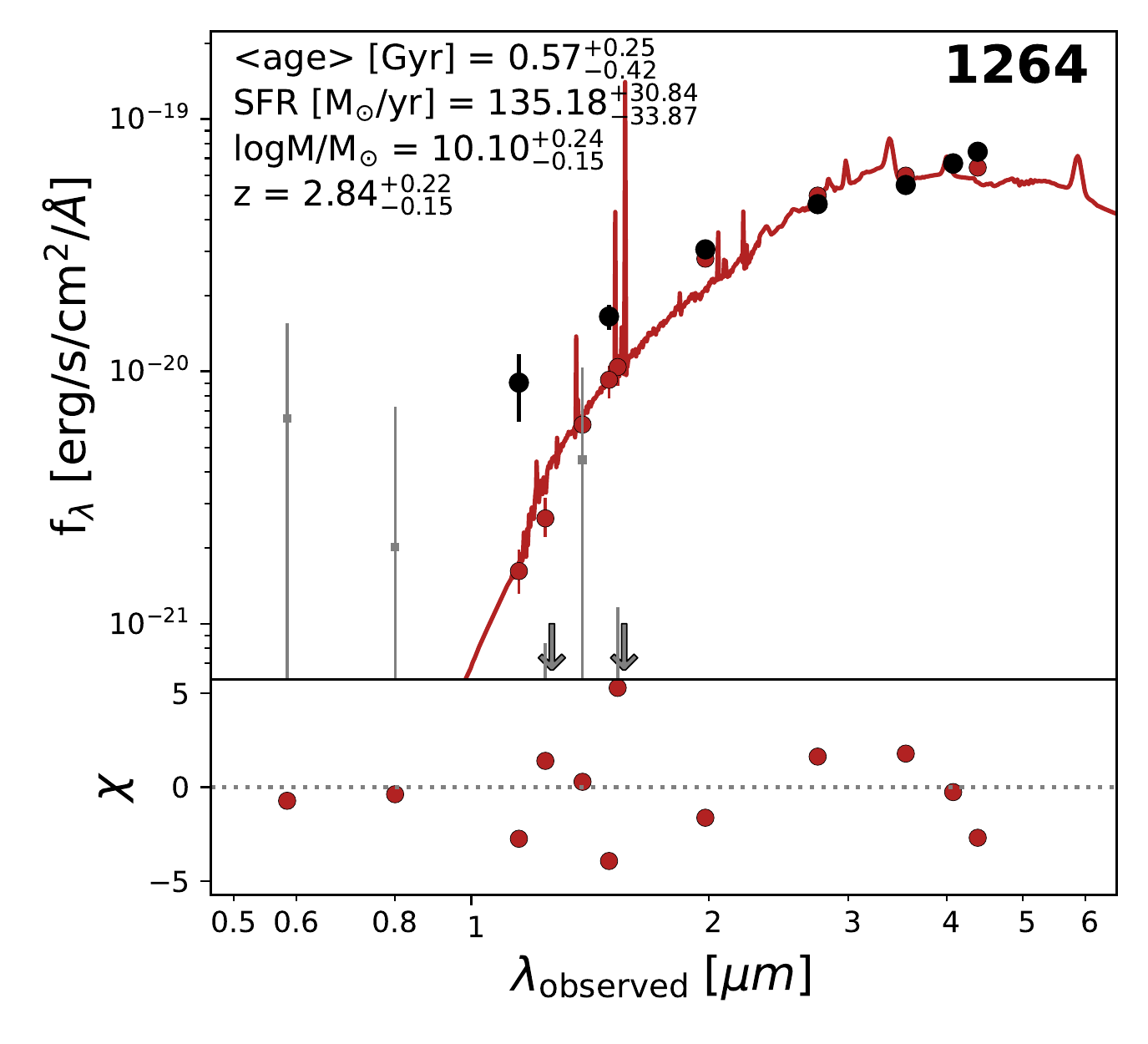}
\includegraphics[ clip,width=0.3\textwidth]{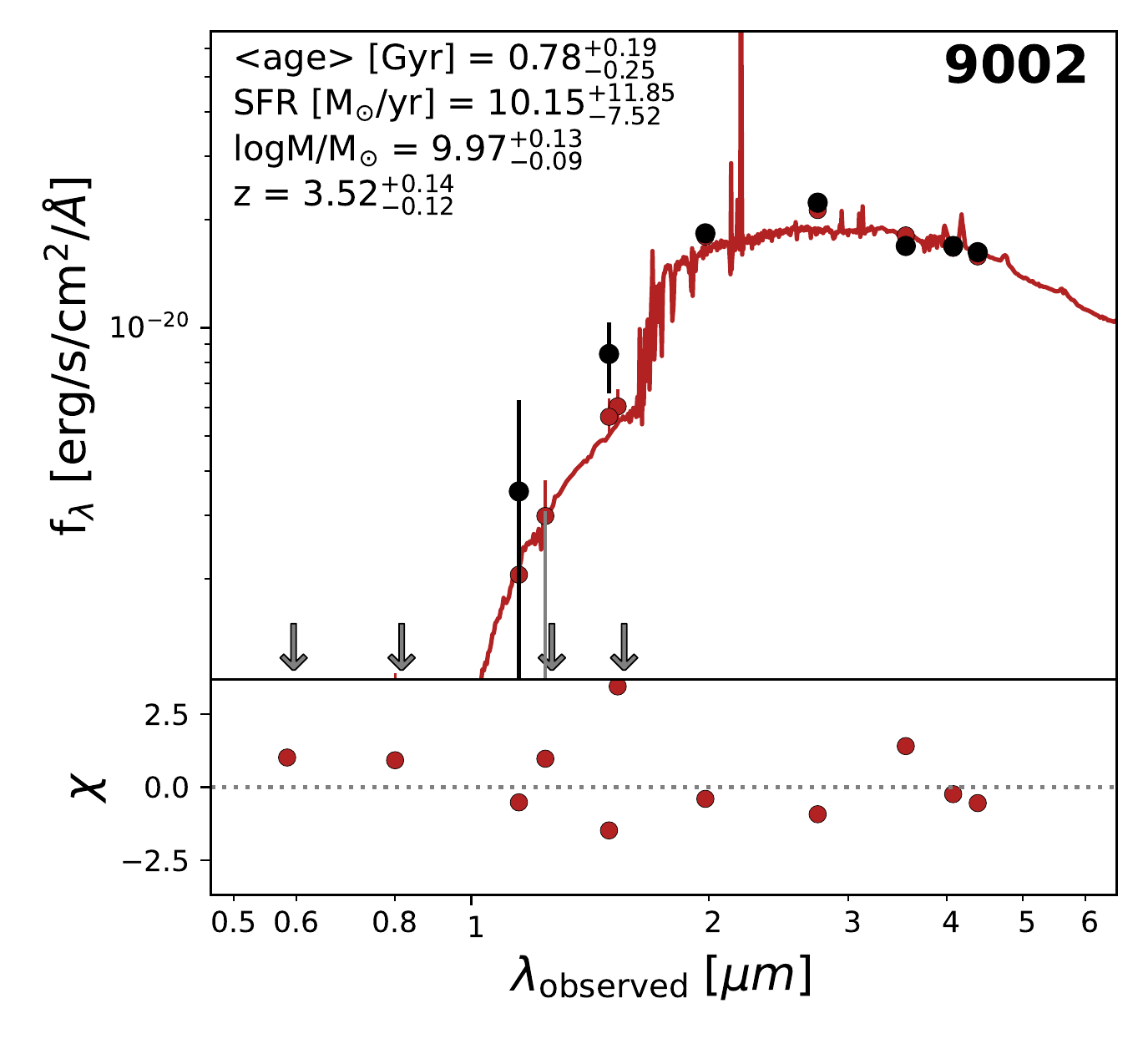}
\includegraphics[ clip,width=0.3\textwidth]{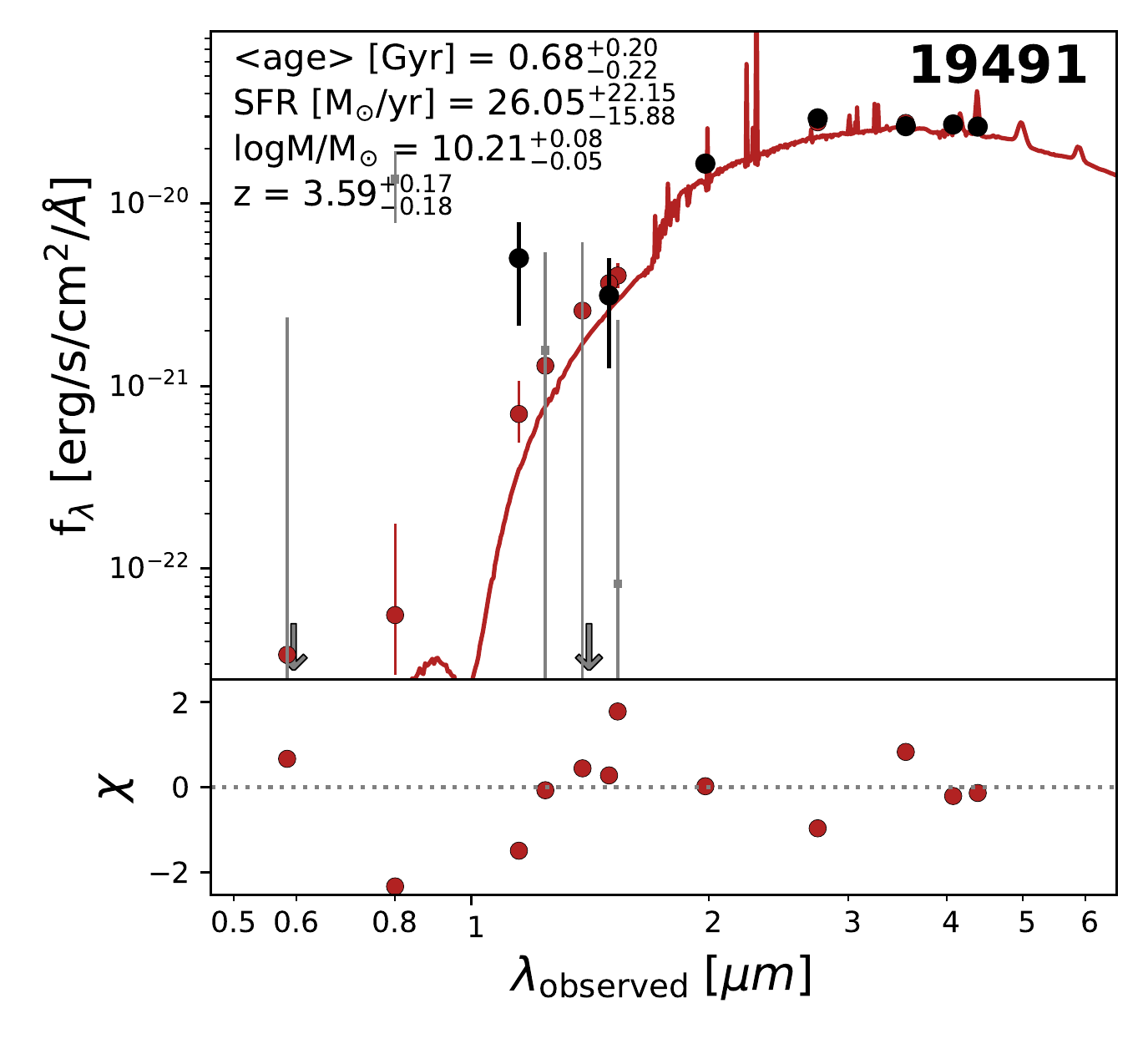}
\includegraphics[ clip,width=0.3\textwidth]{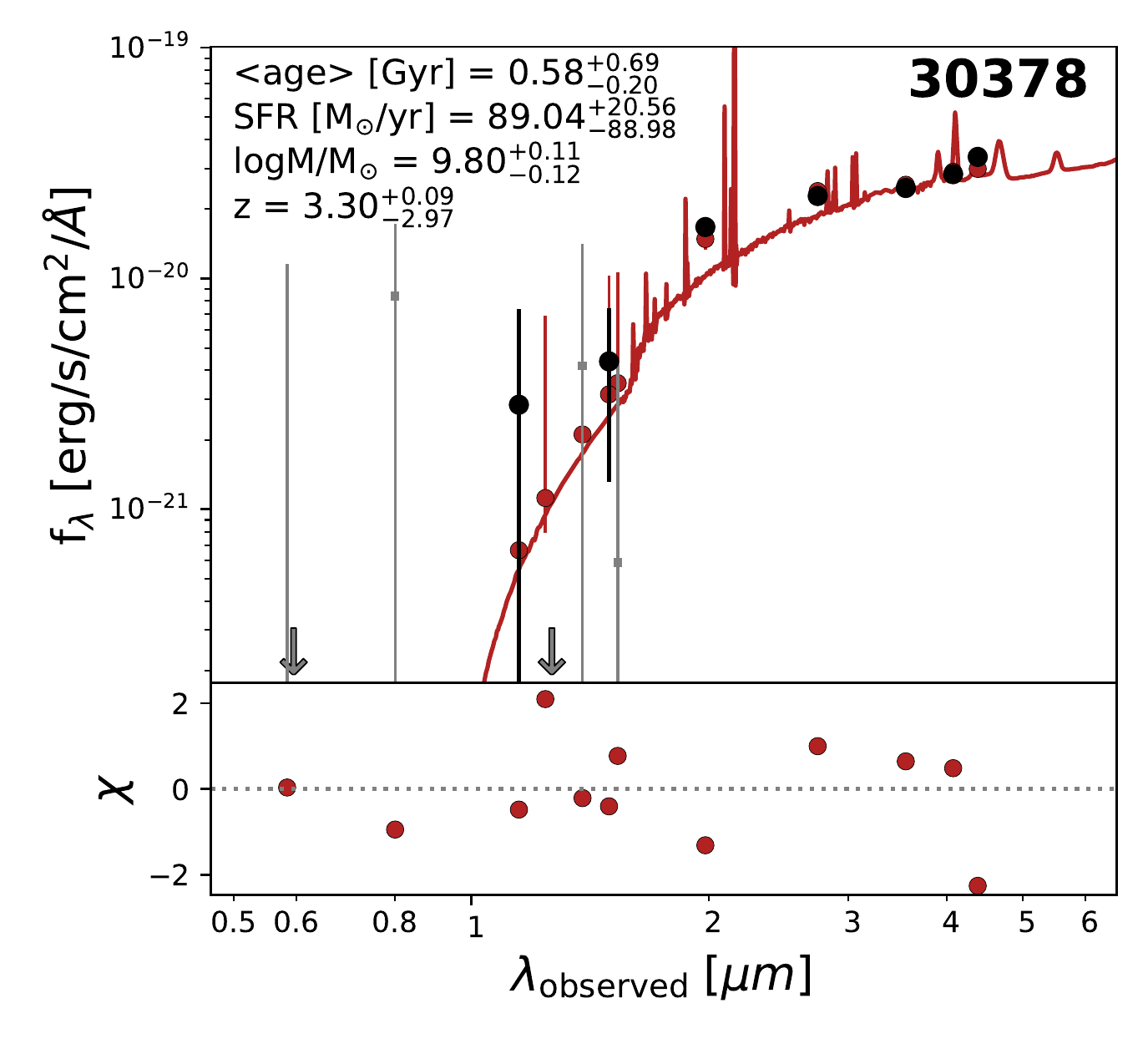}
\includegraphics[ clip,width=0.3\textwidth]{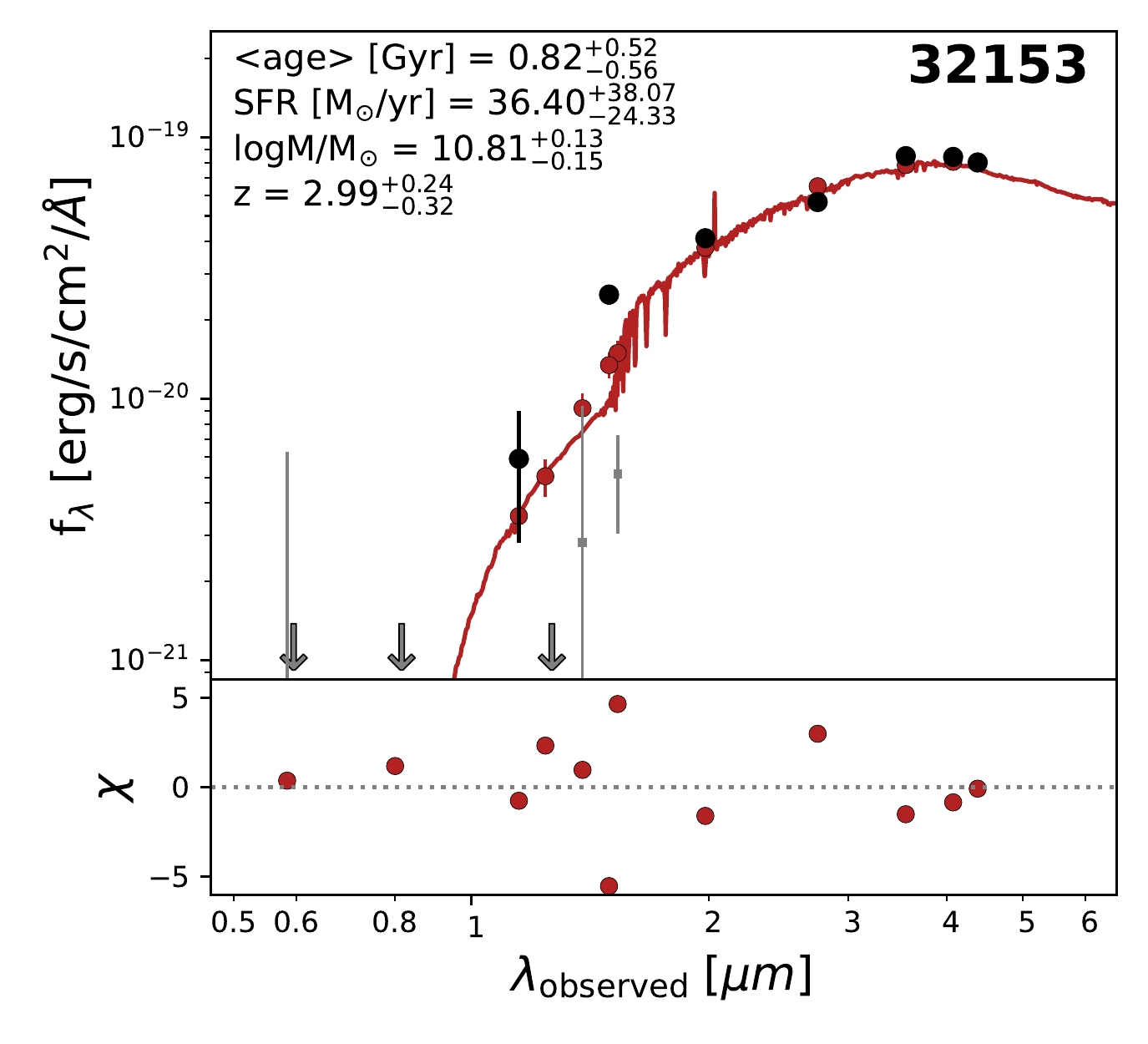}
\includegraphics[ clip,width=0.3\textwidth]{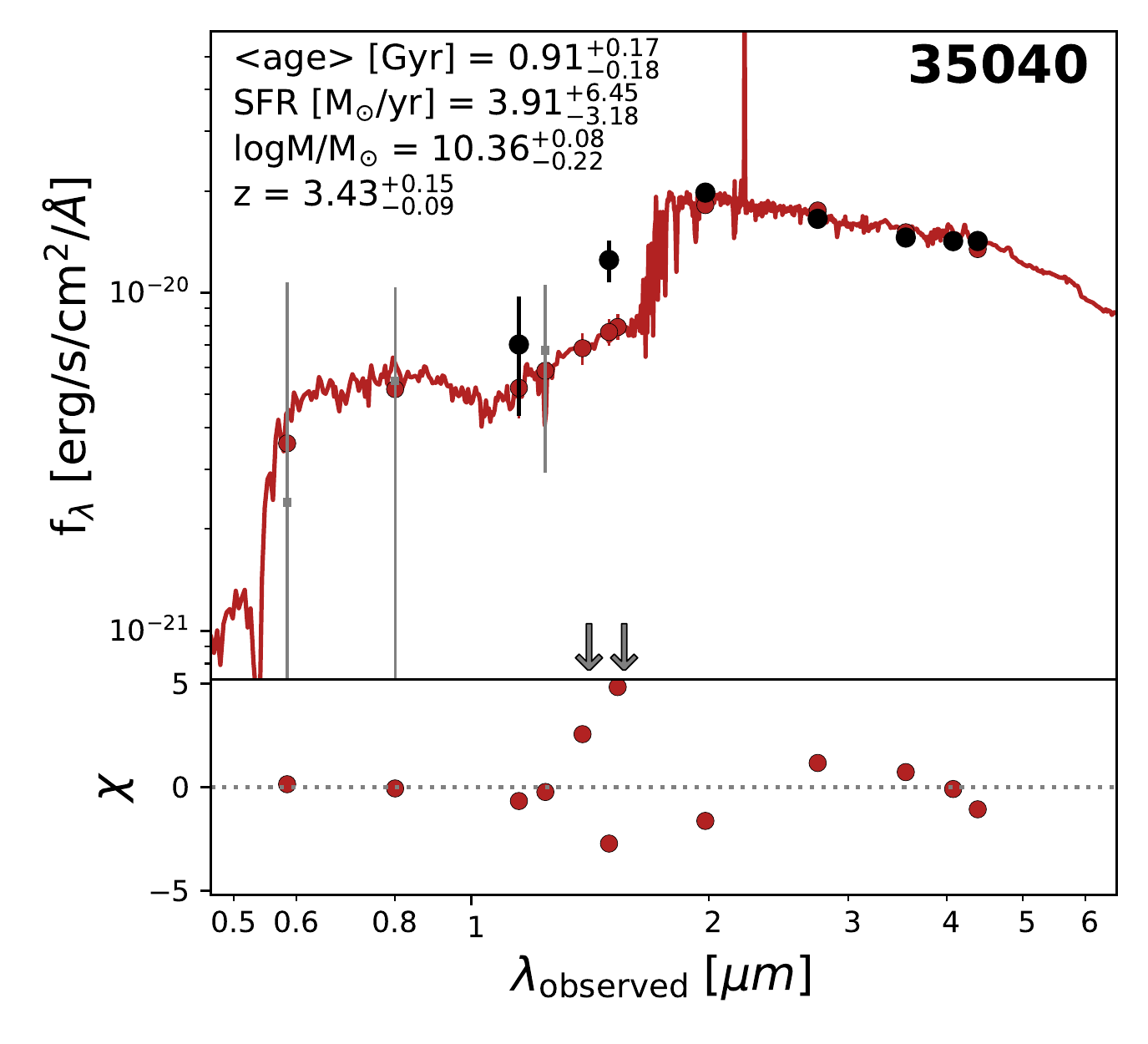}
\includegraphics[ clip,width=0.3\textwidth]{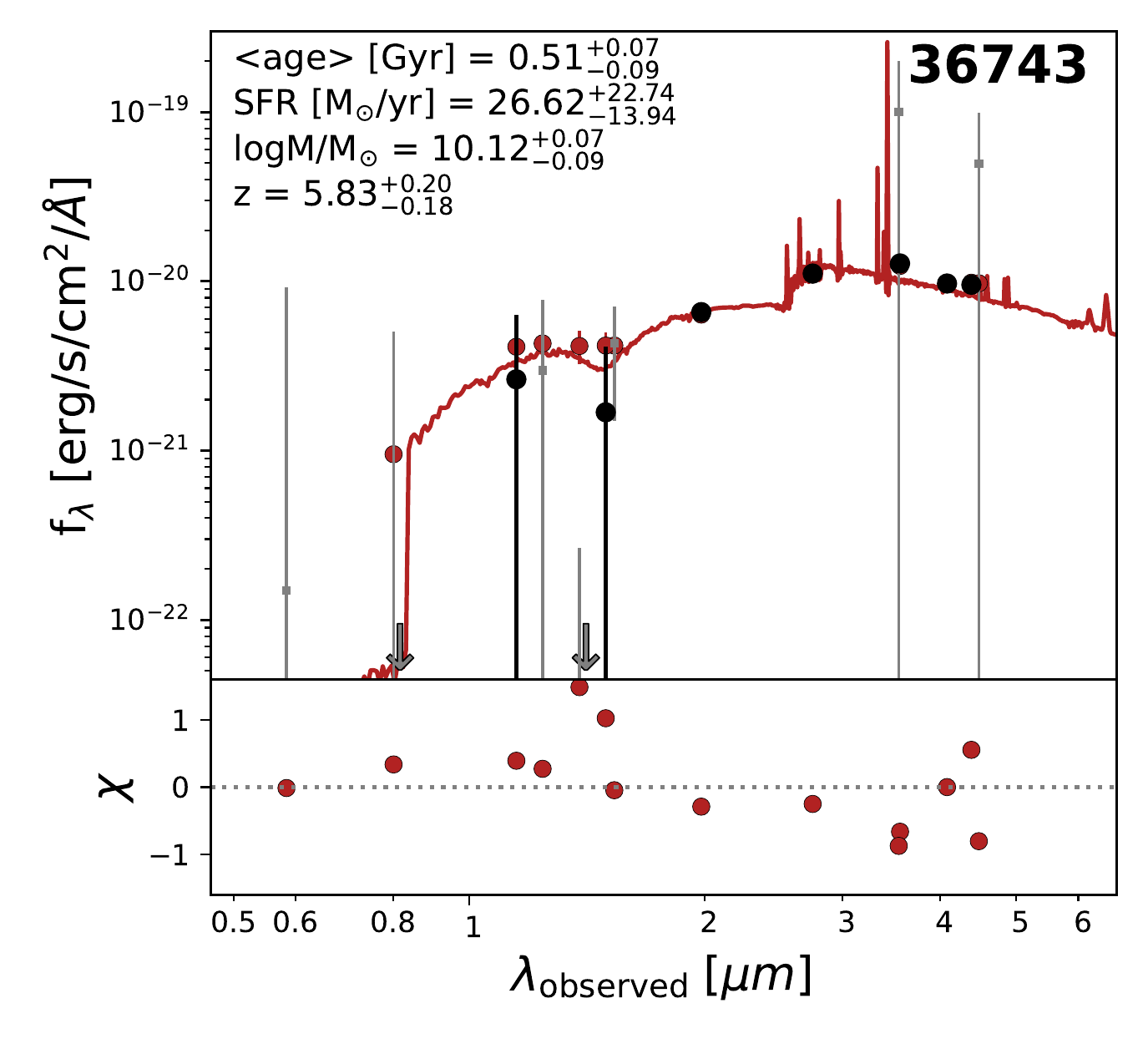}
\includegraphics[ clip,width=0.3\textwidth]{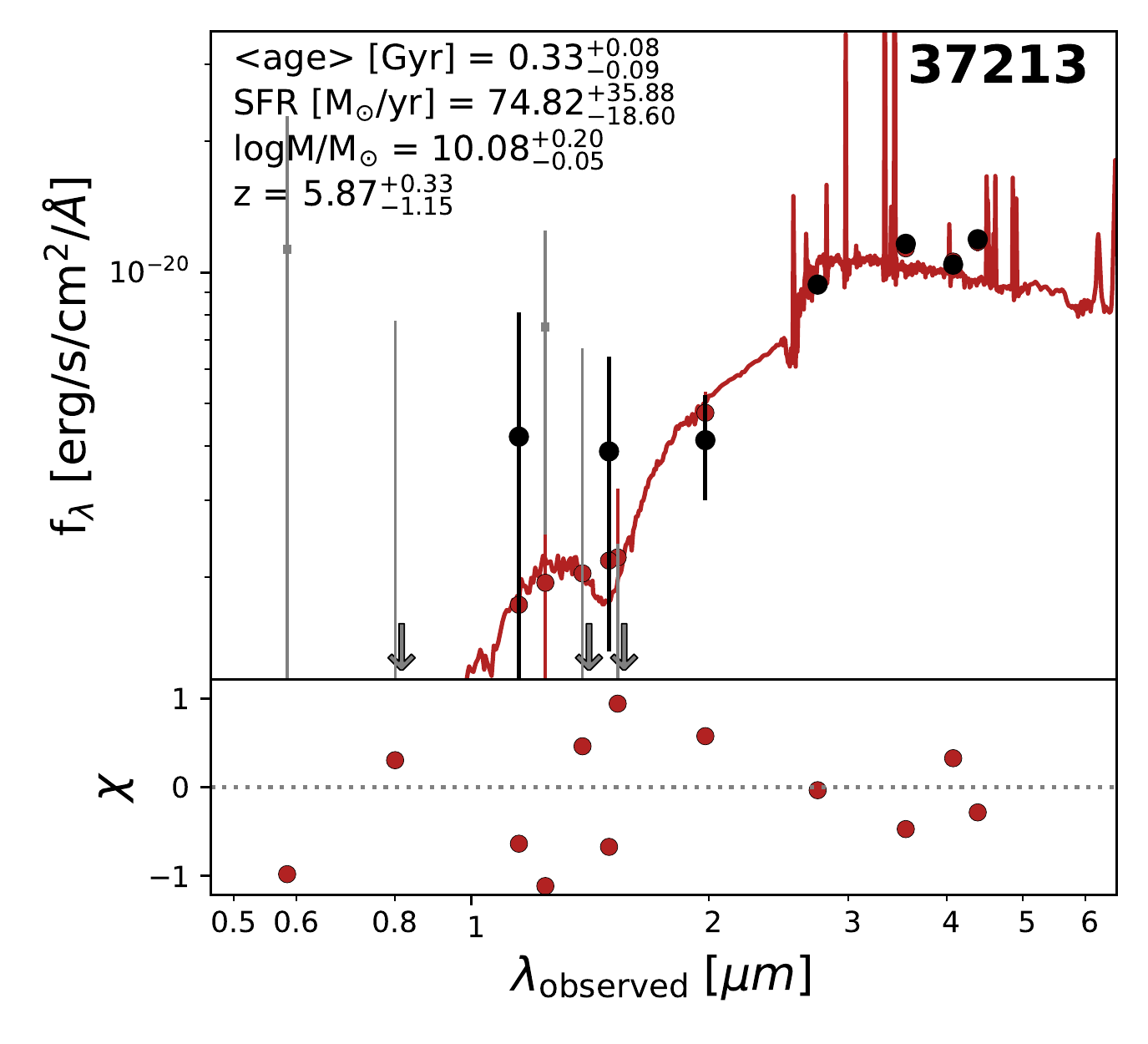}
\includegraphics[ clip,width=0.3\textwidth]{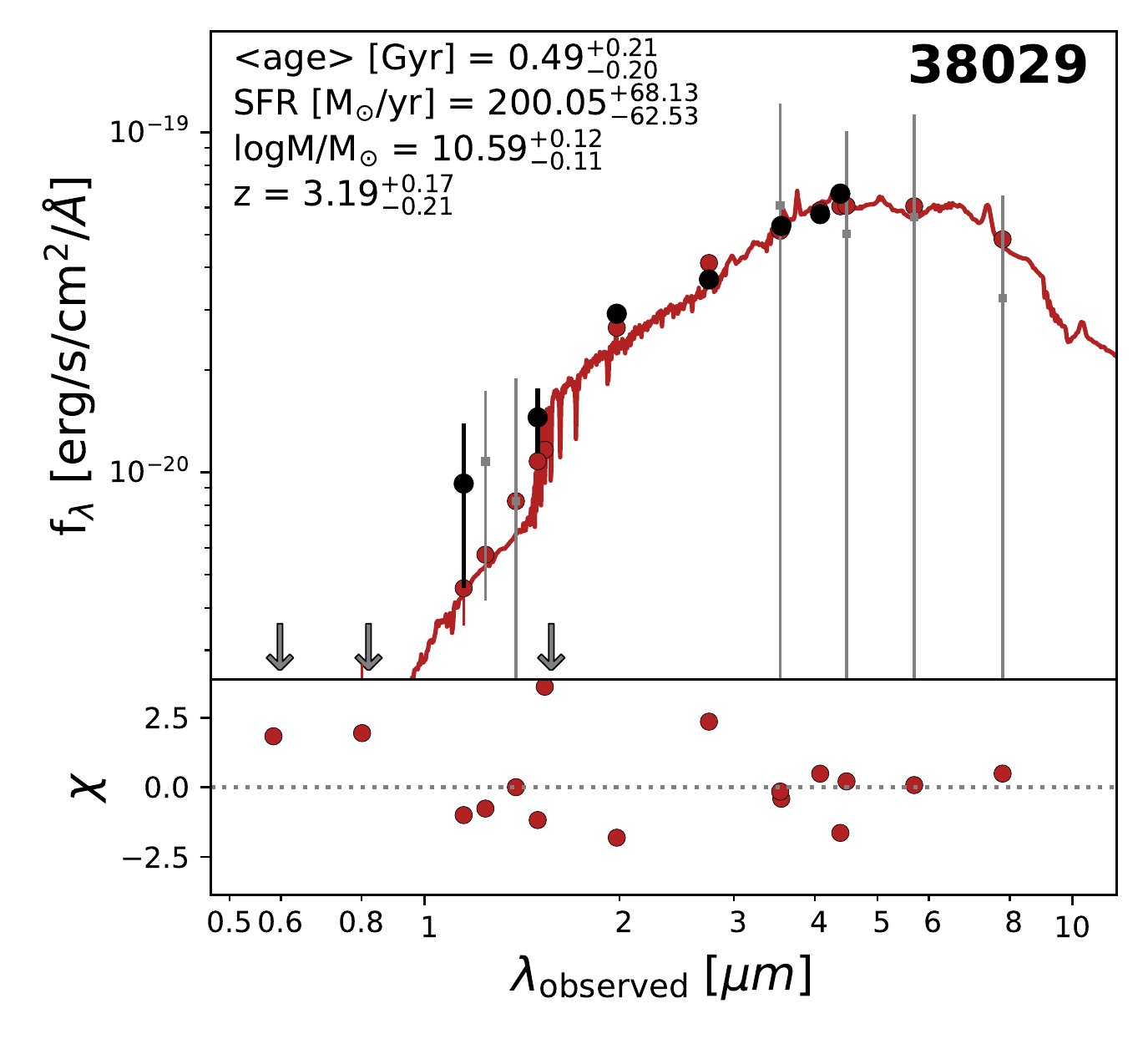}
\includegraphics[ clip,width=0.3\textwidth]{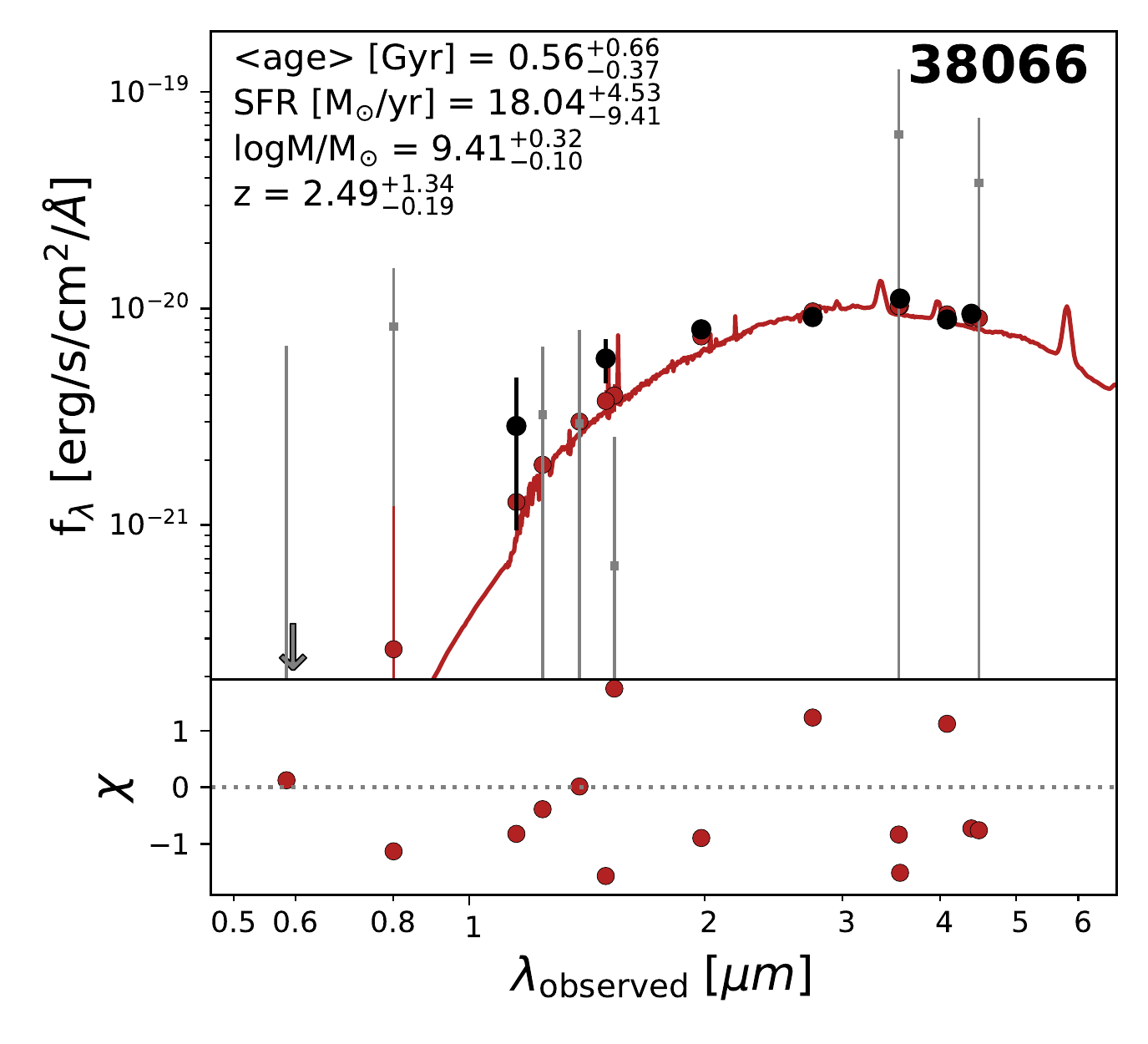}
\includegraphics[ clip,width=0.3\textwidth]{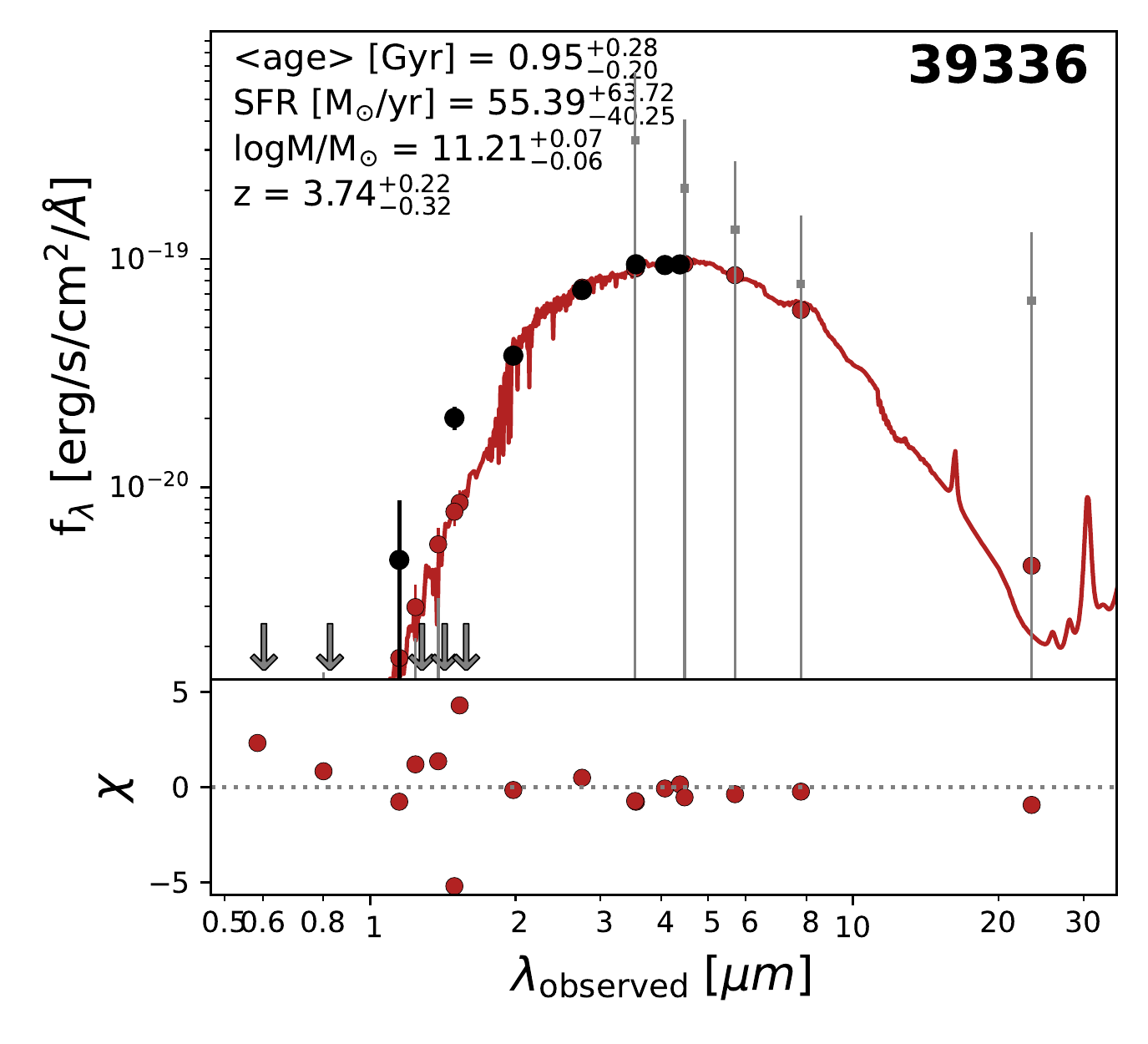}
    \caption{Spectral energy distributions (SEDs) spanning from 0.4 to (in some cases) 24 \micron. Black points show observed \JWST photometry, while grey points show \HST and \emph{Spitzer} photometry. Red shows the best fit model from \parrot. In each case, the SEDs are very red, attributable to a combination of redshift, dust, and age.}
    \label{fig:seds}
\end{figure*}


\section{Redshifts \& Stellar Populations}
\label{subsec:ReSt}
We fit stellar population parameters using the \prospector\  Bayesian inference framework \citep{Johnson:21}, adopting the \prospector-$\alpha$ physical model \citep{leja:17,leja:19}. This models the star formation history by fitting for the mass formed in 7 logarithmically-spaced time bins, assuming a continuity prior which weighs against large changes in star formation rate between bins. The stellar components are modeled with the MIST isochrones \citep{Choi:16,Dotter:16} in FSPS \citep{conroy:10}, a Chabrier IMF, and a two-component dust model. The two-component dust model is particularly important for these red objects: it follows \citet{charlot00} in modeling dust obscuration with a separate birth-cloud and diffuse screen, with the birth-cloud screen affecting only nebular emission and stars with age $<10$ Myr, with up to $A_V \sim 4$ allowed in each component. There is also a flexible attenuation curve following the \citet{noll09} prescription. Nebular emission is powered by the stellar ionizing continuum from the model \citep{Byler:17}. The fitting is sped up by a factor of $\sim$300 using a neural net emulator which mimics stellar population synthesis models, dubbed \parrot\ (get it?) (\citealt{alsing:20}; Mathews et al. in prep).
The parameters we report here are the median of the marginalized posterior probability function; the $1\sigma$ error bars are reported as the 84th--50th and 50th--16th interquartile ranges. We enforce a maximum signal-to-noise ratio of 20 in the observed photometry to allow for both unavoidable model-level errors and potential calibration uncertainties in the new \JWST photometry.

Additionally, in order to avoid spurious high-mass, high-redshift solutions for these very red objects, we adopt a mass function prior for stellar mass, P(logM$^*|$z) (described in Wang et al. in prep). This is constructed by using the observed \citet{leja20} mass functions between $0.2 < z < 3$, inferred using the same \prospector-$\alpha$ stellar populations model. For $z < 0.2$ and $z > 3$, we adopt the nearest-neighbor solution, i.e. the $z=0.2$ and $z=3$ mass functions, as we do not yet have reliable high-resolution rest-frame optical selected mass functions at $z > 3$. In particular this choice allots a conservatively high probability for yet-to-be-discovered populations of high-mass, high-redshift galaxies (hints of which have already been observed, see \citealt{labbe:22}). We allow a solution of $0 < z < 12$ for all objects except 38029, which is allotted a reduced range of $0 < z < 5$ for consistency with the \eazy\ solution (below).

We cross-check our redshifts and stellar masses by also fitting their photometry with the \texttt{\eazy-py} public photometric redshift code \citep{brammer:10}. \eazy\ fits the observed spectral energy distribution (SED) with a set of templates, with the template weight and redshift as free parameters. Because the templates, generated with FSPS \citep{conroy:10}, have associated stellar population properties like age, dust, star formation history, and $M_*/L$ ratio, \eazy\ can also be used to give an estimate of these properties. As a starting point, the default template set \texttt{tweak\_fsps\_QSF\_12\_v3} is used with one modification: an additional \texttt{FSPS} template is generated with an age of $1$Gyr and $A_V=6$ attenuation to allow for more flexibility in reproducing extreme SED shapes.  

Comparing redshifts from \eazy\ and \parrot\ shows a median offset of only $z_{\eazy} - z_{\parrot} = 0.04$, i.e. no significant systematic offsets in either direction. However, the median absolute offset is much larger: $|z_{\eazy} - z_{\parrot}| = 0.4$. This suggests that the photometric redshifts of these objects are not terribly well constrained with existing data owing to their very red and often featureless SEDs. The inclusion of MIRI photometry and/or NIRSpec spectroscopy, when available, will hopefully improve the situation significantly.  The stellar masses have a systematic offset of $\mathrm{log M}_*({\eazy}) - \mathrm{log M}_*({\parrot}) = 0.25$ and a median absolute difference of $0.4$ dex.

The best fit spectral energy distributions and observed photometry of our sample of F444W-selected galaxies 
are shown in Fig.~\ref{fig:seds}. 
In general these are massive, dust-obscured, star-forming galaxies at a median redshift of $z=3.5$ and span the range from $2.4<z<6.5$.
They have a median log stellar mass of $\log M_{\star}/M_{\odot} =$ 10.5 from \eazy\ and 10.1 from \parrot\ with a tail up to very massive galaxies with log(\mstar)=11.2. 
With median dust attenuation values of $A_V\sim2$ and star formation rates of SFR$\sim40 M_\odot$/yr these are dusty star forming galaxies as determined by their rest-optical and near-infrared light, but not extremely so, although we note these quantities are not well-constrained. These are likely an extreme extension (in color and redshift) of the three red spirals found in the SMACS ERO observations by \citet{fudamoto:22}


\begin{figure*}
    \centering
    \includegraphics[width=.3\textwidth]{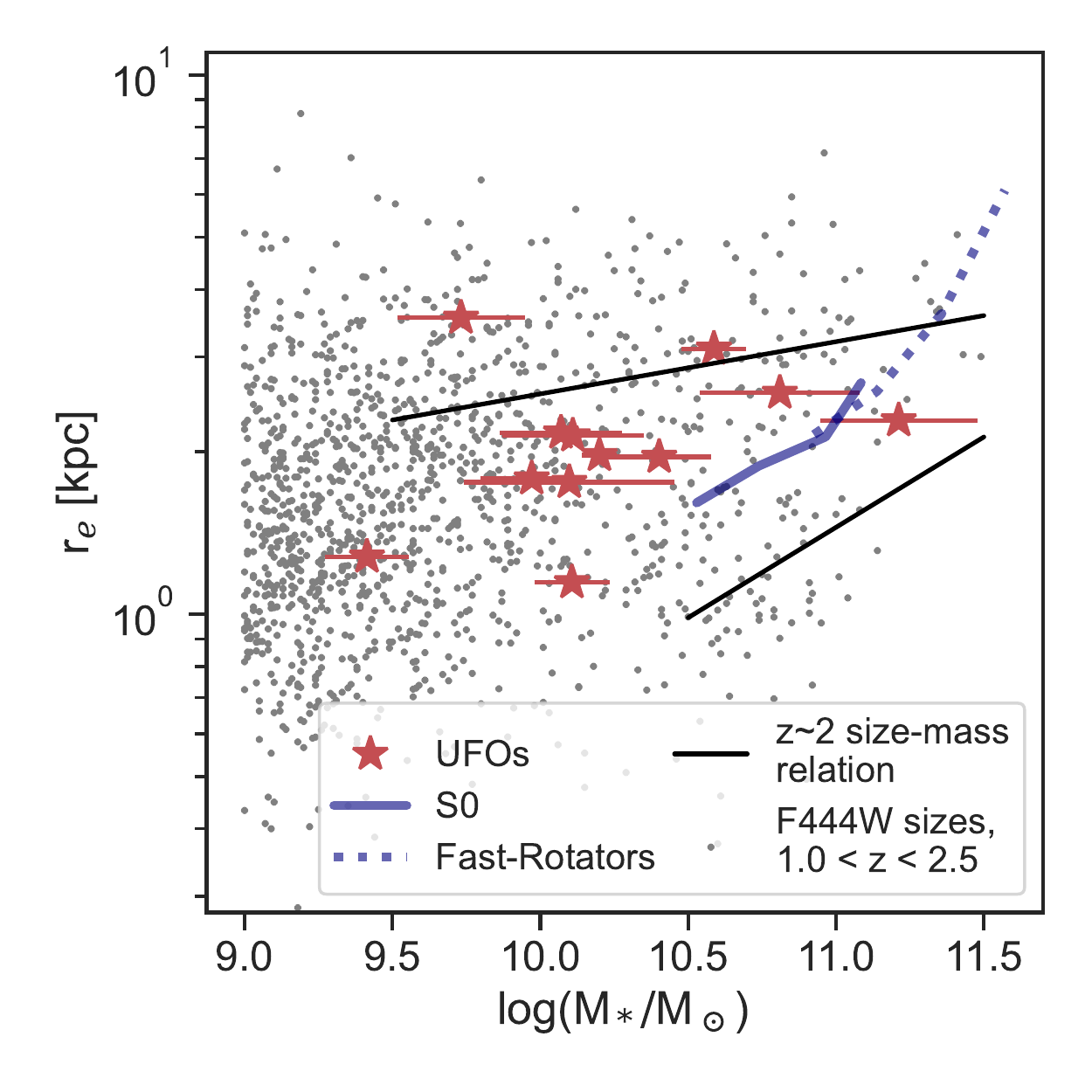}
    \includegraphics[width=.3\textwidth]{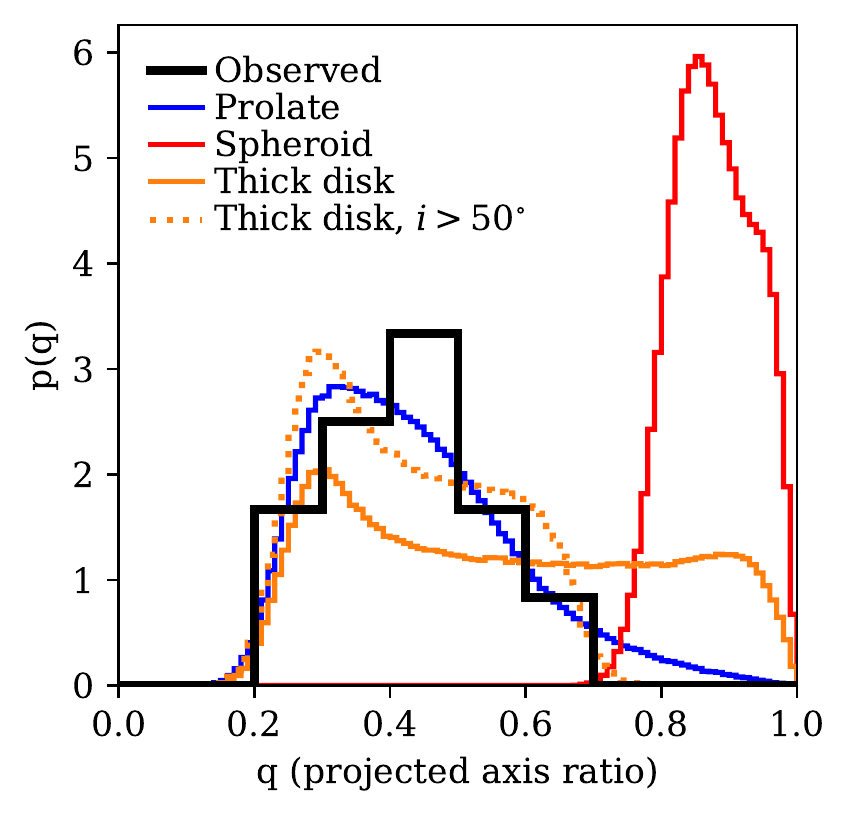}
    \includegraphics[width=.3\textwidth]{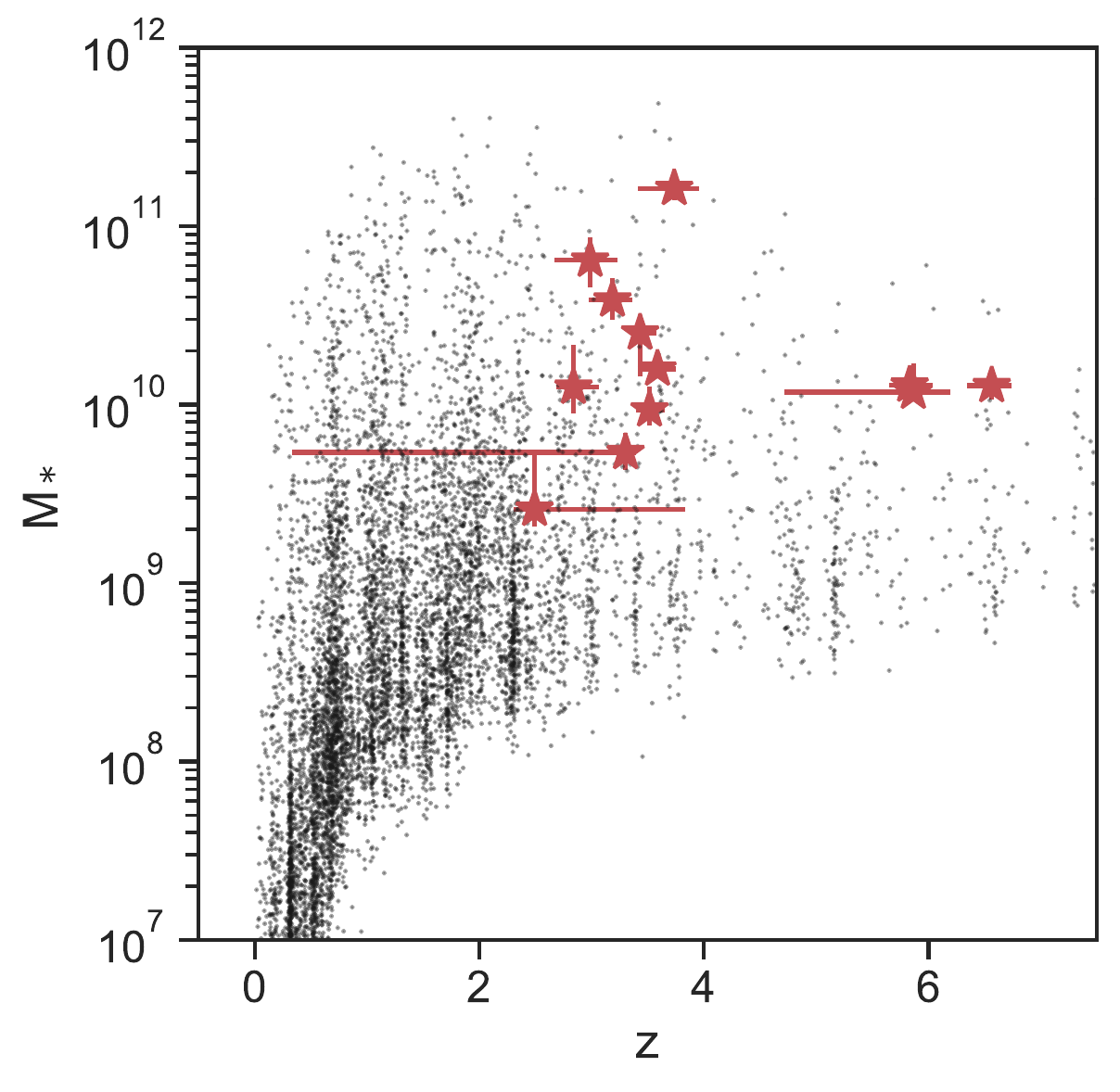}
    \caption{\textbf{Physical Properties.} Left: effective (half-light) semi-major radius in F444W versus stellar mass. The grey points show the F444W sizes from \citet{suess22c} for $1.0<z<2.5$ galaxies at $M_*>10^9M_\odot$, and the red symbols show UFOs. To guide the eye, the black lines show the relationship between half-mass radius and stellar mass at $z\sim2$ \citep{suess:19}, and the blue lines show the size-mass relation for S0s and fast rotators from \citet{bernardi:22}. 
    Center: distribution of observed axis ratios and example population models with various intrinsic 3D shapes. Assuming axisymmetry, the \redrockets\ could have intrinsic disk geometries, with the peaked distribution at $q\sim0.5$ caused by a preferential edge-on orientation for these objects. Without this assumption, prolate shapes also provide a reasonable fit to the data. 
    Right: redshifts and stellar masses for the galaxies in our sample in red. Our galaxies have $2<z<6$ and \mstar$\sim10^{10}-10^{11}~M_\odot$.}
    \label{fig:structpar}
\end{figure*}

\section{Resolved galaxy properties}
\label{subsec:sizes}
We use the \texttt{GALFIT} software package \citep{peng:02, peng:10} to fit the sizes and shapes of galaxies accounting for the point spread function (PSF), following the procedure described in \citet{suess22c}. While the use of stars in the observed images to create empirical PSFs would be preferable, the centers of most stars in our CEERS mosaic are saturated or masked out, necessitating the use of a theoretical PSF. We generate theoretical PSFs with 9$\times$ oversampling on the same 0.04'' pixel scale as our image mosaic using the WebbPSF software \citep{perrin14}. We then rotate these theoretical PSFs to match the position angle of the observations, convolve the 9$\times$ oversampled PSFs with a 9$\times$9 square kernel, and downsample them to the pixel scale of the mosaic. The purpose of conducting the bulk of this procedure with an oversampled PSF image is to minimize distortions from the rotation algorithm. 

For the sake of global warming, \texttt{GALFIT} must be fed images that do not contain more than a handful of sources it must model simultaneously. It also needs a sufficiently large postage stamp to contain all of the light from the galaxy of interest. As a balance, we cut 80$\times$80 pixel postage stamps of our galaxies out of their mosaics, corresponding to $>6r_e$ of all galaxies in our sample. We create a segmentation map of each postage stamp to identify all sources to be modeled or masked. Galaxies that have centers within 3'' of the target galaxy center and are less than 2.5~mag fainter are modeled simultaneously. Fainter and more distant galaxies are masked. With all sufficiently bright galaxies identified, we estimate and subtract the background in each stamp using the SExtractor background algorithm as implemented in \texttt{photutils}. We run \texttt{GALFIT} on the F444W images, fitting S\'ersic profiles to each unmasked galaxy. 

\subsection{Sizes}
Fig.~\ref{fig:structpar}a shows the (semi-major) effective radii at 4.4\micron\ of these galaxies versus their stellar mass. These galaxies have angular sizes of 0.17-0.33" (the lower bound is introduced by selection), which is fairly typical for their fluxes (see Fig.~\ref{fig:selection_v2}b). Physically, they have sizes of 1 to 3~kpc. 
The black lines indicate the mass-weighted size-mass relation at $z\sim2.25$ from \citet[][which builds on the light-weighted size-mass relations from \citealt{mowla:19} and \citealt{vanderwel:14a}]{suess:19}. These half-mass sizes have been shown to be consistent with sizes measured from F444W imaging. These galaxies lie just below the relation for star-forming galaxies. Given that galaxy sizes are expected to grow with redshift \citep[e.g.,][]{mo:98}, the slight offset could easily be attributed to the higher median redshift of this sample. They are more extended than quiescent galaxies at $z\sim2.25$, however when compared to mass-weighted sizes of local early type galaxies (S0s and fast-rotators) from MANGA \cite{bernardi:22}, their sizes could be consistent with the low-mass extension of the $z\sim0$ relation. 


With the same amount of dust, more compact galaxies will have higher dust column densities and hence more dust reddening. Thus, the naive intuition about galaxies which are bright in 4.4\micron, but so reddened as to be undetected with \HST may have been that they would be more compact than the population writ large \citep[e.g.][]{nelson:14}. However, as can be seen in both Fig.~\ref{fig:gallery} and Fig.~\ref{fig:structpar}, these galaxies are typical for their stellar masses. Interestingly, these galaxies, have similar sizes to the FIR sizes of massive galaxies at $z\sim2.5$ \citep{hodge:16,tadaki:20}.


\subsection{Axis ratios}
Fig.~\ref{fig:structpar}b shows the distribution of projected axis ratios $q=b/a$ for the galaxies in our sample (black histogram), along with representative models of projected ellipsoids. Interestingly, all galaxies are significantly flat in projection, with axis ratios $0.24<q<0.65$. This distribution is firmly inconsistent with very round oblate spheroids (red line). 
Absent other effects (i.e., dust), one would expect randomly oriented 3D disks to exhibit flat distributions (orange solid line), not peaked at low $q$. 
This might suggest a likely selection bias against face-on galaxies (orange dashed line). If dust is well-mixed, then our color selection might preferentially select edge-on disks with longer total optical path through the galaxy \citep[e.g.,][]{maller09, wild:11,patel:12,mowla:19}, which could result in a preferentially peaked axis ratio distribution as opposed to the expected $\sim$flat distribution in axis ratio for disky systems. In other words, face-on systems are more likely to be brighter at bluer wavelengths and hence may be excluded from our sample.

Another possibility is that the low-$q$ values reflect a prolate population (blue line). Although such a model could also be consistent with the observed distribution, this would be at odds with the results for rest-frame optical axis ratio distributions of high mass galaxies at slightly lower redshifts (e.g., \citealt{vanderwel:14b}, \citealt{zhang:19}, \citealt{zhang:22}).  The S\'ersic indices of these systems are for the most part close to $n=1$, which could be consistent with a disk- or prolate-dominated population in F444W. 


Although the intrinsic 3D shapes of galaxies cannot be measured directly, they can be statistically inferred based on the projected axis ratio distribution through statistical modeling  \citep[e.g.][]{vanderwel:14b, zhang:19}. 
We follow the modeling procedure of \citet{chang:13} and \citet{vanderwel:14b}, 
adopting the median axis ratio uncertainty 
$\delta q$
for our sample and using dynamic nested sampling \citep[\texttt{Dynesty,}][]{speagle:20} to determine the posterior distribution  of the intrinsic galaxy geometries (i.e., $E=1-(C/A)$ and  $T=[1-(B/A)^2]/[1-(C/A)^2]$ with intrinsic axes lengths $A\geq B\geq C$, as in \cite{vanderwel:14b}). 
For a free fit of $\mu_E, \sigma_E, \mu _T, \sigma_T$, the fact that the observed distribution peaks at 
$q\sim0.5$ 
is best described by population with prolate geometries \citep[see, e.g.][]{vanderwel:14b, zhang:22}. Alternatively, if we assume these galaxies are axisymmetric ($T=0$), then the individual objects all have high probabilities ($P(\mathrm{disk})\gtrsim 50\%$) of being disks (defined as  $C/A\leq0.4$). Such an axisymmetric disk geometry would be consistent with the possibility that these objects are dusty, highly-inclined disks, which is what one might guess based on the images in Fig.~\ref{fig:gallery}.
Ultimately, this statistical fitting is inherently uncertain as the sample size is small and potentially dramatically biased due to the size and color-selections.

Interestingly, submillimeter-selected galaxy samples appear to have similar axis ratio distributions, suggesting significant population overlap. These results have led to interpretations that those galaxies are either disk \citep{hodge:16} or triaxial systems \citep{gullberg:19}.  These previous \emph{HST} studies of sub-mm bright galaxies have suffered from similar small number statistics to definitively characterize the 3D structures of heavily dust-obscured systems. However, further \emph{JWST} imaging should continue to robustly detect starlight from dusty star forming galaxies, ultimately building up much larger samples that will be better suited to answering these statistical questions.

\begin{figure}
    \centering
    \includegraphics[width=0.5\textwidth]{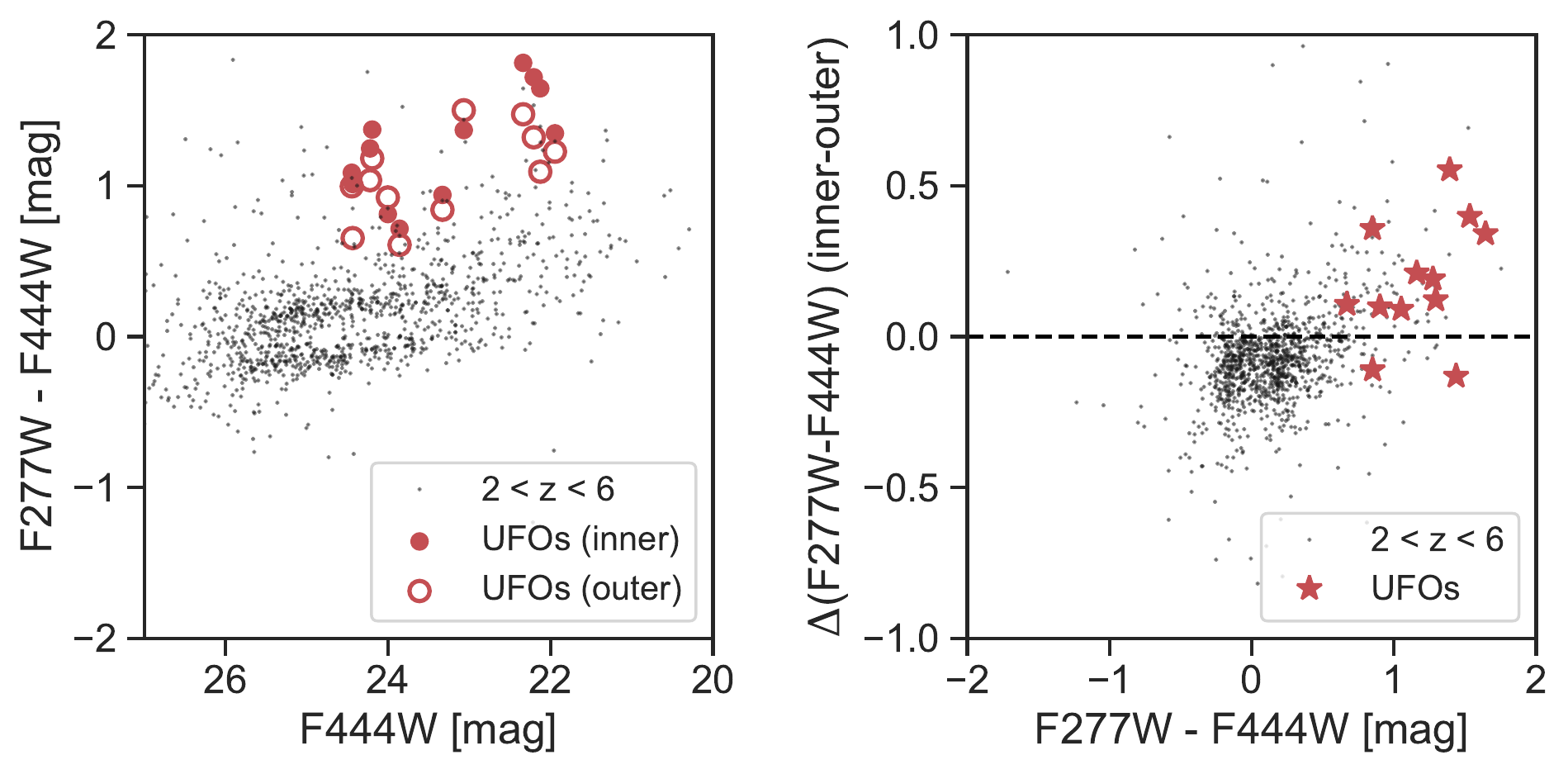}
    \caption{\textbf{Color gradients (or lack thereof).} Left: comparison of the F277W-F444W color in the central 0.3" (filled circles) and a 0.3-0.5" annulus (open circles) as a function of total F444W magnitude. Grey points show the colors of all galaxies with EAZY redshifts $2 < z< 6$. Right: difference between the inner and outer F277W-F444W color as a function of the total F277W-F444W color. Not only are the UFOs significantly redder than most galaxies at similar redshifts, they are red throughout their disks. }
    \label{fig:colorgradients}
\end{figure}

\subsection{Color gradients (or lack thereof)}
\label{sec:color}
One of the most surprising and striking features of these galaxies is their seeming lack of strong color gradients: they appear to be red throughout their disks. We quantify this by comparing their colors at different radii. To do this, we perform photometry with both 0.32" and 0.5" apertures. We then compare the color in the central 0.32" and the 0.32-0.5" annulus colors derived by subtracting the 0.32" from the 0.5" aperture fluxes. As shown in Fig.~\ref{fig:colorgradients}, we find fairly small color gradients: the 0.32" and 0.5-0.32" colors are similar, and significantly redder than integrated galaxy colors of the full sample (black points). Two galaxies in our UFO sample have slightly redder centers than outskirts, while the remaining ten are red throughout or have slightly bluer centers. The central aperture has a median color $F277W-F444W = 1.30$ mag while the outer annulus has 1.06 mag and the median color gradient is 0.16~mag. 
The lack of strong negative color gradients in these systems is surprising and in contrast with a canonical view of massive disk galaxies with red bulges and blue disks. 

As shown in the right panel of Fig.~\ref{fig:colorgradients}, the color gradients in our sample significantly differ from those of typical galaxies, which tend to have redder centers and bluer outskirts. Previous work has suggested that massive galaxies at all epochs exhibit significant color gradients due to a combination of age, dust, and metallicity differences \citep[e.g.,][]{suess:19,miller:22,suess:22}. This difference may suggest that previous studies based on \emph{HST} imaging would have excluded a fully optically-thick population included in this work.




\section{Discussion and Conclusions}
\label{sec:conc}
\emph{JWST} is already allowing us to see the universe with new eyes, providing unprecedented spatial resolution in the infrared. With the \emph{JWST}/NIRCam imaging released from the early CEERS program, we investigate the stellar structures of extremely red galaxies to which \emph{HST} was previously blind, but are bright at 4\micron. This population includes 12 physically extended galaxies at $2\lesssim z\lesssim6$. Although photometric redshifts and stellar population modeling are still preliminary, these galaxies are massive ($M_*\sim10^{10}-10^{11}$) and are a mix of dusty and star forming with some existing older stellar populations. 
This sample highlights the fact that the \emph{JWST} discovery space extends studies of galaxy stellar structures to later cosmic epochs during which we thought we had a reasonable census of the universe already.

Perhaps the most noteworthy result stems from the flattened shapes of these \HST-dark galaxies. These massive, star-forming galaxies are the likely progenitors of today’s massive galaxies, which tend to be bulge/spheroid-dominated. Disk-dominated objects at these cosmic times are expected to be gas-rich and gravitationally unstable, resulting in clump formation and migration to galactic centers, building central bulges \citep[e.g.,][]{ceverino:17}. Finally, studies of the kinematics of star-forming galaxies show that they become progressively more dispersion-dominated and less rotation-dominated at earlier cosmic times \citep{Gnerucci:11,Kassin:12,Wisnioski:15,Wisnioski:19,price:20}. Taken together, the expectation may have been that the stellar bodies of these objects would already host significant bulges. This, however, is not what we observe in this sample. These objects have uniformly low axis ratios ($q  \lesssim 0.6$) and S\'ersic indices close to one, inconsistent with being significantly bulge-dominated/oblate at 4.4\,\micron. If we assume axisymmetry, these galaxies are likely to be disks. If not, the lack of high axis ratio objects could potentially suggest a prolate population. We think the former is more likely given our sample selection. Selecting things which are extended and with little to no blue light may result in not selecting face-on disks, which are less obscured than edge-on disks \citep[e.g.,][]{wild:11, patel:12, mowla:19}. If we loosen our color cut slightly, we do indeed see high axis ratio objects with similar masses and sizes. 

Another surprising aspect of these objects is the spatial patterns in their colors. As a result of building up from the inside-out \citep[e.g.][]{nelson:12}, we may expect age and hence color gradients in extended massive galaxies \citep[e.g.,][]{suess:19,mosleh:20,miller:22}. We also expect the centers of galaxies to be more dust-obscured than large radii, also contributing to color gradients \citep[e.g.][]{Nelson:16a,Tacchella:18}.  
Indeed, \cite{suess22c} find smaller sizes in 4.4\,\micron\ than 1.5\,\micron, implying light that is more concentrated in the red than the blue, consistent with the physical picture of red bulges within blue disks. Hence, it came as a surprise that most of these objects are red throughout their disks. While two show some blue low-surface brightness light at large radii, the bulk are uniformly red throughout or even slightly redder in their outskirts. This suggests significant dust attenuation through the galaxies, not merely in their centers. 
The color gradients in our sample are consistent with the idea of looking through large dust columns at all radii in edge-on disks. A more in-depth discussion for the physical drivers of color gradients (or lack thereof) will be presented in T. Miller et al. in prep. 

Although this Letter focuses on a small sample of galaxies, this study emphasizes that JWST presents a new opportunity to connect the known populations of heavily dust-obscured star-forming galaxies that dominate the star formation history of the Universe at cosmic noon \citep{casey:14,zavala:21} with stellar mass-selected samples of galaxies across cosmic time. It will allow us to connect the distribution of existing stellar mass (e.g., with NIRCam at 4\,\micron\ in this study) with dust-obscured star formation (e.g., with MIRI). The full picture will extend early studies of the morphological evolution in massive galaxies \citep[e.g.,][]{hodge:20}. Furthermore, the less stochastically growing stellar components will likely help to test expectations of progenitor-descendant matching \citep[e.g.,][]{toft:14}, in which uncertainties are currently dominated by the timescales of the sub-mm-bright phase.

We have mostly compared this sample of galaxies to massive disk galaxies with growing bulges; hoewver it is important to note that these galaxies could be the early progenitors of a very different population of quiescent galaxies in the local Universe. Many quiescent, massive galaxies are either extremely flattened (lenticular) or significantly rotationally supported (fast-rotators;  \citealt{cappellari:11}). We refer back to the the mass-weighted size-mass relation for S0 and fast-rotating galaxies from the MANGA survey \citep{bernardi:22}, as shown in Fig.~\ref{fig:structpar}a. This small sample is insufficent to test whether continued star-formation and self-similar growth, as implied by the uniform colors in this sample, would evolve the population onto the local relation. However, it is notable that, unlike the full population of star-forming galaxies at cosmic noon, this sample does not require significant structural evolution to reside within existing local descendants. Further studies with more precise redshifts and number densities will be necessary to perform a more careful progenitor-descendant linking.

Finally, understanding the true nature of these ultra-red galaxies will require more precise measurements of their redshifts, which will dramatically improve our ability to characterize their stellar populations. In this context, this avenue of study will benefit greatly from the spectroscopic capabilities of \emph{JWST}. Furthermore, later this year, the CEERS program will cover the NIRCam footprint with MIRI imaging, which will further pin down the long-wavelength emission from these galaxies, breaking the degeneracy between age, dust, and redshift and tightening the confidence intervals.





    


\begin{table*}
\centering
\caption{\HST-dark Galaxy Sample and Derived Properties}
\label{tab:sample}
\renewcommand{\arraystretch}{1.3}
\begin{tabular}{lcccccccccc} 
 \hline
ID  &  RA   &  DEC   & F115W & F150W & F444W & $r_e$ & $z_{\eazy}$ &  $z_{\parrot}$ & $\log(M_{*,\eazy})$ & $\log(M_{*,\parrot})$ \\
    &   [deg]   &  [deg]   &  [mag]  &  [mag] & [mag] & ["] &  &  & [$M_\odot$]  &  [$M_\odot$]  \\[0.5ex] 
 \hline\hline
       375 & 214.84026471 &  52.80111372 &  28.19 &  26.69 &  24.22 &   0.21 & 6.6 & $6.6^{0.2}_{0.2}$ &  10.53 & $10.1^{0.1}_{0.1}$ \\ 
      1264 & 214.80997035 &  52.80974637 &  27.40 &  26.18 &  22.21 &   0.23 & 3.2 & $2.8^{0.2}_{0.2}$ &  10.50 & $10.1^{0.2}_{0.1}$ \\ 
      9002 & 214.87066655 &  52.84610577 &  28.43 &  26.90 &  23.86 &   0.24 & 3.5 & $3.5^{0.1}_{0.1}$ &  10.32 & $10.0^{0.1}_{0.1}$ \\ 
     19491 & 214.92575659 &  52.91852695 &  28.04 &  27.98 &  23.33 &   0.27 & 3.9 & $3.6^{0.2}_{0.2}$ &  11.01 & $10.2^{0.1}_{0.0}$ \\ 
     30378 & 214.76796482 &  52.81635558 &  28.66 &  27.62 &  23.07 &   0.50 & 4.0 & $3.3^{0.1}_{0.1}$ &  10.45 & $ 9.7^{0.1}_{0.1}$ \\ 
     32153 & 214.90114122 &  52.83810115 &  27.87 &  25.73 &  22.13 &   0.31 & 2.5 & $3.0^{0.2}_{0.2}$ &  10.58 & $10.8^{0.1}_{0.2}$ \\ 
     35040 & 214.86578588 &  52.88342028 &  27.67 &  26.48 &  24.00 &   0.25 & 3.1 & $3.4^{0.1}_{0.1}$ &   9.91 & $10.4^{0.1}_{0.2}$ \\ 
     36743 & 214.93158625 &  52.92100264 &  28.74 &  28.65 &  24.44 &   0.32 & 4.7 & $5.8^{0.2}_{0.2}$ &  10.27 & $10.1^{0.1}_{0.1}$ \\ 
     37213 & 214.89183472 &  52.93389900 &  28.23 &  27.75 &  24.19 &   0.37 & 6.0 & $5.9^{0.3}_{0.3}$ &  10.54 & $10.1^{0.2}_{0.0}$ \\ 
     38029 & 214.97747231 &  52.95348812 &  27.37 &  26.32 &  22.34 &   0.40 & 3.1 & $3.2^{0.2}_{0.2}$ &  10.44 & $10.6^{0.1}_{0.1}$ \\ 
     38066 & 214.92577657 &  52.95444502 &  28.65 &  27.30 &  24.45 &   0.16 & 2.5 & $2.5^{1.3}_{1.3}$ &   9.38 & $ 9.4^{0.3}_{0.1}$ \\ 
     39336 & 215.02153788 &  52.99130129 &  28.09 &  25.96 &  21.95 &   0.27 & 2.0 & $3.7^{0.2}_{0.2}$ &  10.51 & $11.2^{0.1}_{0.1}$ \\ 
\hline 
\end{tabular}
\end{table*}

\begin{acknowledgements}
EJN acknowledges support from HST-AR-16146. KAS acknowledges the UCSC Chancellor’s Postdoctoral Fellowship Program for support. The Cosmic Dawn Center (DAWN) is funded by the Danish National Research Foundation under grant No. 140. Cloud-based data processing and file storage for this work is provided by the AWS Cloud Credits for Research program.  RB acknowledges support from the Research Corporation for Scientific Advancement (RCSA) Cottrell Scholar Award ID No: 27587. H\"U gratefully acknowledges support by the Isaac Newton Trust and by the Kavli Foundation through a Newton-Kavli Junior Fellowship. This work is based in part on observations made with the NASA/ESA/CSA James Webb Space Telescope. The data were obtained from the Mikulski Archive for Space Telescopes at the Space Telescope Science Institute, which is operated by the Association of Universities for Research in Astronomy, Inc., under NASA contract NAS 5-03127 for JWST. These observations are associated with program DD-ERS\#1345 (PI: Finkelstein).
\end{acknowledgements}

\vspace{5mm}
\facilities{\emph{JWST}(NIRCam), \emph{HST}(WFC3)}

\software{
\texttt{numpy} \citep{harris:20}, 
\texttt{scipy} \citep{virtanen:20}, 
\texttt{matplotlib} \citep{Hunter:2007}, 
\texttt{astropy} \citep{astropy:13,astropy:18}, 
\texttt{grizli} \citep{Brammer:21}, \texttt{photutils} \citep{Bradley:20}, 
\texttt{astrodrizzle} \citep{avila:15}, 
\texttt{GALFIT} \citep{peng:02,peng:10}, 
\texttt{Dynesty} \citep{speagle:20}, 
\texttt{\prospector} \citep{Johnson:21}
}

\bibliographystyle{aasjournal}

\bibliography{all}

\end{document}